\documentclass[aps,prl,showpacs,twocolumn,superscriptaddress,nofootinbib]{revtex4-1}

\usepackage{graphicx}
\usepackage{color}      
\usepackage{bm}
\usepackage{amsmath}
\usepackage{times, txfonts}

\newcommand{\mX}{\mathcal{X}}
\newcommand{\mY}{\mathcal{Y}}

\newcommand{\nmax}{N_{\mbox{\footnotesize{max}}}}

\makeatletter
\def\simleq{\mathrel{\mathpalette\gl@align<}}
\def\simgeq{\mathrel{\mathpalette\gl@align>}}
\def\gl@align#1#2{\lower.6ex\vbox{\baselineskip\z@skip\lineskip\z@
     \ialign{$\m@th#1\hfill##\hfil$\crcr#2\crcr\sim\crcr}}}
\makeatother

\begin{document}

\title{Doubly magic nuclei from Lattice QCD forces at $M_{PS}=$~469~MeV/c$^2$}

\author{C. McIlroy}
\email{c.s.mcilroy@surrey.ac.uk}
\affiliation{Department of Physics, University of Surrey, Guildford GU2 7XH, UK}

\author{C. Barbieri}
\email{C.Barbieri@surrey.ac.uk}
\affiliation{Department of Physics, University of Surrey, Guildford GU2 7XH, UK}

\author{T. Inoue}
\affiliation{Nihon University, College of Bioresource Sciences, Kanagawa 252-0880, Japan}

\author{T. Doi}
\affiliation{Theoretical Research Division, Nishina Center, RIKEN, Wako 351-0198, Japan}
\affiliation{ iTHEMS Program and iTHES Research Group, RIKEN, Wako 351-0198, Japan}

\author{T. Hatsuda}
\affiliation{Theoretical Research Division, Nishina Center, RIKEN, Wako 351-0198, Japan}
\affiliation{ iTHEMS Program and iTHES Research Group, RIKEN, Wako 351-0198, Japan}

\date{\today}

\begin{abstract}
We perform {\em ab~initio} self-consistent Green's function calculations of the closed shell nuclei $^{\rm 4}$He, $^{\rm 16}$O and $^{\rm 40}$Ca, based on two-nucleon potentials derived from Lattice QCD simulations, in the flavor SU(3) limit and at the pseudo-scalar meson mass of 469~MeV/c$^{\rm 2}$. 
The  nucleon-nucleon interaction is obtained using the HAL QCD method and its   short-distance repulsion is treated by means of ladder resummations outside the model space.  Our results show that this approach diagonalises ultraviolet degrees of freedom correctly. Therefore, ground state energies can be obtained from infrared extrapolations even for the relatively hard potentials of HAL QCD. 
  Comparing to previous Brueckner Hartree-Fock calculations, the total binding energies are sensibly improved by the full account of many-body correlations.  The results suggest an interesting possible behaviour in which nuclei are unbound at very large pion masses and islands of stability appear at first around the traditional  doubly-magic numbers when the pion mass is lowered toward its physical value.
The calculated one-nucleon spectral distributions are qualitatively close to those of real nuclei even for the pseudo-scalar meson mass considered here.

\end{abstract}

\maketitle

{\it Introduction}.
Quantum chromodynamics (QCD) is expected to ultimately explain the structure 
 and the interactions of all hadronic systems, together with  small corrections of electroweak origin.
At the moment,  systematic and non-perturbative calculations of QCD can 
be carried out only by Lattice QCD (LQCD). Indeed, 
 high-precision studies have been shown to be possible e.g., in the case of single-hadron 
 masses~\cite{Borsanyi:2014jba}. 
``Direct'' calculations of multi-baryon systems
have also been attempted on a lattice by several groups~\cite{Beane:2011iw,Yamazaki:2012hi,Yamazaki:2015asa,Orginos:2015aya,Berkowitz:2015eaa}.  However, the typical excitation energy $\Delta E$ for multi-baryons is  one to two orders of magnitude smaller than ${\cal O}(\Lambda_{\rm QCD})$.
Accordingly, high statistic data with very large Euclidean times $t \simgeq \hbar/\Delta E \sim$ 10-100 fm/c is required.
This  is still far beyond reach due to exponentially  increasing errors in $t$ and $A$ (the atomic number),
 as demonstrated theoretically and numerically in recent studies  
 ~\cite{Iritani:2016jie,Aoki:2016dmo,Iritani:2016xmx}.

In this work, we follow a different route and perform a first {\it ab~initio} study
for medium mass atomic nuclei directly based on QCD by taking a two-step strategy. 
In the first step, 
we extract the nuclear force from {\it ab~initio} LQCD calculations 
with the  HAL~QCD method. This procedure generates consistent two-, 
three- and many-nucleon forces in a systematic way~\cite{Ishii:2006ec,Aoki:2009ji,Aoki:2012tk,Aoki2013PRD}.
Note that the HAL~QCD interactions  obtained directly from spatiotemporal hadron-hadron correlations on the lattice 
are faithful to the scattering phase shifts and binding-energies  by construction.
This is done systematically without the fitting procedures needed by phenomenological potentials. 
Furthermore, explorative studies
of three-nucleon potential show that this is weaker than the corresponding nucleon-nucleon (NN)
force, in agreement with the empirically observed hierarchy of nuclear forces~\cite{Doi:2011gq}. Thus, 
the interaction can be applied to larger nuclei.
In the second step, we calculate the properties of nuclei
with {\it ab~initio}  many-body methods using the LQCD potentials as input.
The advantage of the HAL~QCD approach is that one can extract
the potential dictating all the elastic scattering states below the inelastic threshold 
 from the lattice data for $t \simgeq 1$ fm/c~\cite{HALQCD:2012aa}.
This makes the LQCD calculation of potentials affordable with reasonable statistics, 
together with the help of advanced computational algorithms~\cite{Doi:2012xd,Detmold:2012eu,Gunther:2013xj,Nemura2016CPC}.
We can then take the advantage of the 
recent developments in nuclear many-body theories 
to calculate various information on nuclei such as binding energies
and spectral distributions.  Note that a similar  two-step strategy has also been taken in 
Ref.~\cite{Barnea2005lqcd,Contessi2017pilessEFT}  where, however, effective field theories have been used to fit the LQCD data.

Past {LQCD} studies in the {flavor} SU(3) limit {by the  HAL~QCD collaboration}
have led to interactions in both the nucleon and hyperon sectors with masses of the pseudo-scalar meson (which corresponds to the pion) as low as \hbox{$M_{PS}$$={}$469~MeV/c$^2$.} In these cases, potentials have been obtained for the {$^1$S$_0$} and the coupled $^3$S$_1$-$^3$D$_1$ channels~\cite{Inoue:2010es,Inoue:2011ai}.  
Exploratory calculations 
 based on {these} HAL~QCD  potentials were performed in the  Brueckner Hartree-Fock (BHF) approach~\cite{Inoue:2013nfe,Inoue:2014ipa}. This is  quantitative enough to give the essential underlying physics for infinite matter  but it is less reliable in finite systems.  More sophisticated calculations are needed in order to go beyond the mean-field level, which is necessary to properly predict binding energies and to  describe the truly complex structures of nuclei at low energy. 
BHF becomes even more questionable for finite nuclei due to assumptions  with the unperturbed single particle spectrum where there is a problem in the choice between a continuous or a gap form, neither of which is completely satisfactory.

{\em Ab~initio} theories for medium mass nuclei have advanced greatly in the past decade and methods
 such as coupled cluster and self-consistent Green's function (SCGF) are now routinely employed to study the structure of full isotopic chains up to Ca and Ni~\cite{Soma2014s2n,Hergert2014Ni,Hagen2014CCMrev,Dickhoff2004ppnp}, with the inclusion of three-nucleon forces~\cite{Carbone2013tnf}.
Their use with soft interactions from chiral effective field theory have led to first principle predictions of experimental total binding energies~\cite{Hebeler2015ChiralRev,Cipollone2013prl} and nuclear radii~\cite{Ekstrom2015NNLOsat,Lapoux2016prlOx,Ruiz2016NatCa48} with unprecedented accuracies. 
Interactions like the HAL~QCD potentials pose a bigger challenge for calculations of medium mass isotopes due to their short-distance inter-nucleon repulsion. However, the SCGF method has the advantage that two-nucleon scattering (ladder) diagrams at large momenta---needed to resolve the short-range repulsion---can be dealt with explicitly by solving the Bethe-Goldstone equation (BGE) in the excluded space~\cite{Barbieri2002o16,Barbieri2006plbO16,Barbieri2009Ni56prc}. This route was exploited in the past to study spectral strength distribution but we extend it to binding energies in the present work and find that this is accurate enough to make statements on the performance of the present HAL~QCD potentials.
Thus, this work is also a first step toward advancing many-body approaches that can handle hard interactions for large atomic masses.
This is also important since new LQCD calculations at nearly the physical pion mass are currently in progress ~\cite{Doi:2015oha,Ishii:2016zsf,Sasaki:2016gpc,Nemura:2016sty}.

The BHF study of Ref.~\cite{Inoue:2013nfe} showed that the HAL~QCD interactions in the SU(3) limit do not bind at very large pion masses except for the lowest available value \hbox{$M_{PS}$$={}$469~MeV/c$^2$.} The saturation of nuclear matter in this case is also confirmed by later SCGF calculations~\cite{CarboneHALprivcomm}. 
Thus, this is a suitable  choice to investigate possible self-bound nuclei at large pion masses. In this work, we will focus on this potential and refer to it as the `HAL469$_{\hbox{\small  \em SU(3)}}$' interaction, or `HAL469' for simplicity.


\begin{figure}[t]
\includegraphics[width=0.95\columnwidth]{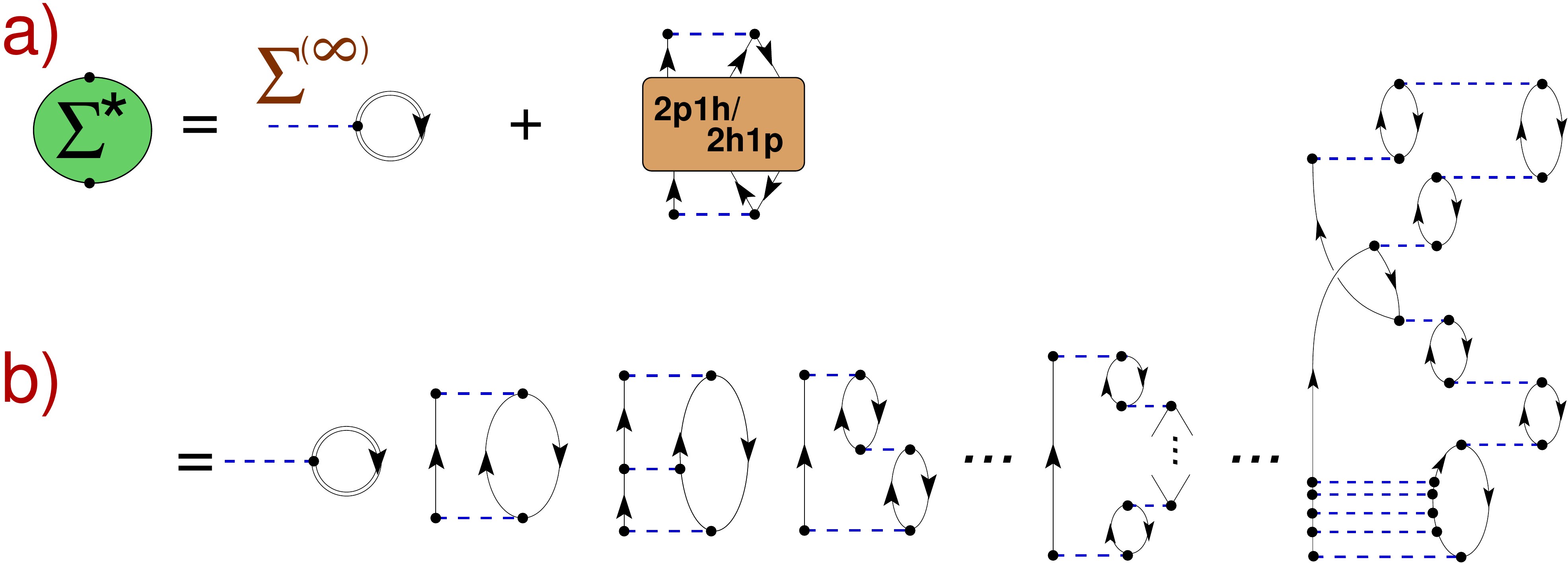}
\caption{(Color online) 
Diagrammatic content of the  ADC(3)  approximation.  {\em (a)}  The self-energy splits in a static mean field part and an energy dependent contribution according to Eq.~\eqref{eq:Sig_split}. {\em (b)} The dynamic contributions are obtained as infinite resummations of ladder (pp/hh)  and the ring (ph) diagrams. The  ADC(3) approach includes static corrections to the coupling of nucleons to intermediate excitations (matrix $\mathbf{D}$ in Eq.~\eqref{eq:Sig_split}), an example of which is shown by the top portion of the last Goldstone diagram~\cite{Barbieri2017LNP}.}
\label{fig:ADC3}
\end{figure}

{\it Formalism}.
We follow closely the approach of Ref.~\cite{Barbieri2009Ni56prc} and focus on the single-particle  propagator given by~\cite{fetter2003quantum,dickhoff2008many}:
\begin{align}
g_{\alpha\beta}(\omega) ={}& \sum_{n} \frac{(\mX_{\alpha}^n)^*\mX_{\beta}^n}{\omega - \varepsilon_n^+ + i\eta} + \sum_{k} \frac{\mY_{\alpha}^k(\mY_{\beta}^k)^*}{\omega - \varepsilon_k^- - i\eta} \; ,
\label{eq:gLeh}
\end{align}
 where $\mX_{\alpha}^{n} \equiv \langle\Psi_{n}^{A+1}|c_{\alpha}^{\dagger}|\Psi_{0}^{A}\rangle$ ($\mY_{\alpha}^k \equiv  \langle\Psi_{k}^{A-1}|c_{\alpha}|\Psi_{0}^{A}\rangle$) are the spectroscopic amplitudes, $\varepsilon_n^+\equiv E_n^{A+1}-E_0^A$ ($\varepsilon_k^-\equiv E_0^A-E_k^{A-1})$ are their  quasiparticle energies and $c_{\alpha}^{\dagger} (c_{\alpha})$ are the second quantisation creation (annihilation) operators. 
In these definitions, $|\Psi_{n}^{A+1}\rangle$ and $|\Psi_{k}^{A-1}\rangle$ represent the exact eigenstates of the $(A \pm 1)$-nucleon system, while $E_{n}^{A+1}$ and $E_{k}^{A-1}$ are the corresponding  energies. 
We perform calculations within a spherical harmonic oscillator model space, indicated as $P$, and use Greek indices,  \hbox{$\alpha$, $\beta$, ...} to label its basis functions. Within this space, the one-body propagator is obtained by solving the Dyson equation
with an irreducible self-energy given by 
\begin{align}
\Sigma_{\alpha\beta}^{\star}(\omega) ={}& \Sigma_{\alpha\beta}^{(\infty)} + \sum_{i j}\textbf{D}_{\alpha i}^{\dagger} \left[\frac{1}{\omega - (\textbf{K} +\textbf{C}) \pm i\eta}\right]_{i j} \textbf{D}_{j \beta} \; .
\label{eq:Sig_split}
\end{align}
This expression is the sum of a mean-field (MF) term, $\Sigma^{(\infty)}$,  
and the contributions from dynamical correlations. The coupling~(\textbf{D}) and interactions matrices~(\textbf{K} and  \textbf{C}) are computed in the third-order algebraic diagrammatic construction [ADC(3)] approximation that generates non-perturbative all-order summations of 2p1h and 2h1p configurations, as shown in Fig.~\ref{fig:ADC3}~\cite{Schirmer1983ADCn,Barbieri2007Atoms,Barbieri2017LNP}.
We follow the \emph{sc0} approximation of Refs.~\cite{Barbieri2009Ni56prc,Soma2014GkvII}, in which $\Sigma^{(\infty)}$ is calculated exactly from the fully-dressed propagator $g(\omega)$  while \textbf{D}, \textbf{K} and  \textbf{C} are written in terms of a simplified MF reference propagator, $g^{REF}(\omega)$, that is chosen to best approximate  $g(\omega)$ through its first two moments at the Fermi energy (see Ref.~\cite{Barbieri2009Ni56prc} for details). 

For forces with a sizeable  short-range repulsion, like the HAL~QCD {interactions},  usual truncations of the oscillator space (of up to 12  shells in this case) are not sufficient and a resummation of ladder diagrams in the excluded Hilbert space, $Q\equiv{\bf 1}-P$, is required. We do this by solving the BGE in $Q$ according to Refs.~\cite{HJORTHJENSEN1995125,engeland2008cens} and add the corresponding diagrams to the MF self-energy, which becomes energy dependent~\cite{Barbieri2009Ni56prc}: 
\begin{align}
\Sigma_{\alpha\beta}^{(\infty)}(\omega) ={}& \sum_{\gamma \, \delta}\int\frac{d\omega'}{2\pi i}T^{BGE}_{\alpha\gamma, \, \beta\delta}(\omega+\omega') \, g_{\gamma \delta}(\omega') \, e^{i \omega' \eta} 
\nonumber \\
={}& \sum_{\gamma \, \delta}\sum_k \,  T^{BGE}_{\alpha\gamma, \, \beta\delta}(\omega + \varepsilon_k^-) \, \mY_{\delta}^k(\mY_{\gamma}^k)^* \; ,
\label{eq:SigMF}
\end{align}  
where $T^{BGE}_{\alpha\gamma, \, \beta\delta}(\omega)$ are the elements of the scattering t-matrix in the excluded space.  We then extract a static effective interaction that we use to calculate the ADC(3) self-energy (the last term of Eq.~\eqref{eq:Sig_split}) within the model space $P$.
To do this, we  solve the Hartree-Fock (HF) equations with the MF potential of Eq.~\eqref{eq:SigMF}:
\begin{align}
\sum_\beta  \left\{ 
  \langle\alpha|\frac{p^2}{2m_N}|\beta\rangle +  \Sigma_{\alpha\beta}^{(\infty)}(\omega = \varepsilon_{r}^{HF})
 \right\} \; \psi^r_\beta 
 ={}&   \varepsilon_{r}^{HF}  \, \psi^r_\alpha \; ,
\label{eq:GHF}
\end{align} 
where latin indices label HF states, 
and define a static interaction in this HF basis similarly to Refs.~\cite{PhysRevC.66.044301,Barbieri2009Ni56prc}:
\begin{align}
V_{r s, p q} ={}& \frac{1}{2}\left[ T^{BGE}_{r s, p q}(\varepsilon_{r}^{HF} + \varepsilon_{s}^{HF}) + T^{BGE}_{r s, p q}(\varepsilon_{p}^{HF} + \varepsilon_{q}^{HF})  \right] \; .
\label{eq:Veff}
\end{align}
The $V_{r s, p q}$ matrix elements are then transformed back to the harmonic oscillator space to be used in the computations.

\begin{figure*}[t]
\includegraphics[height=0.46\columnwidth,width=0.63\columnwidth,clip=true]{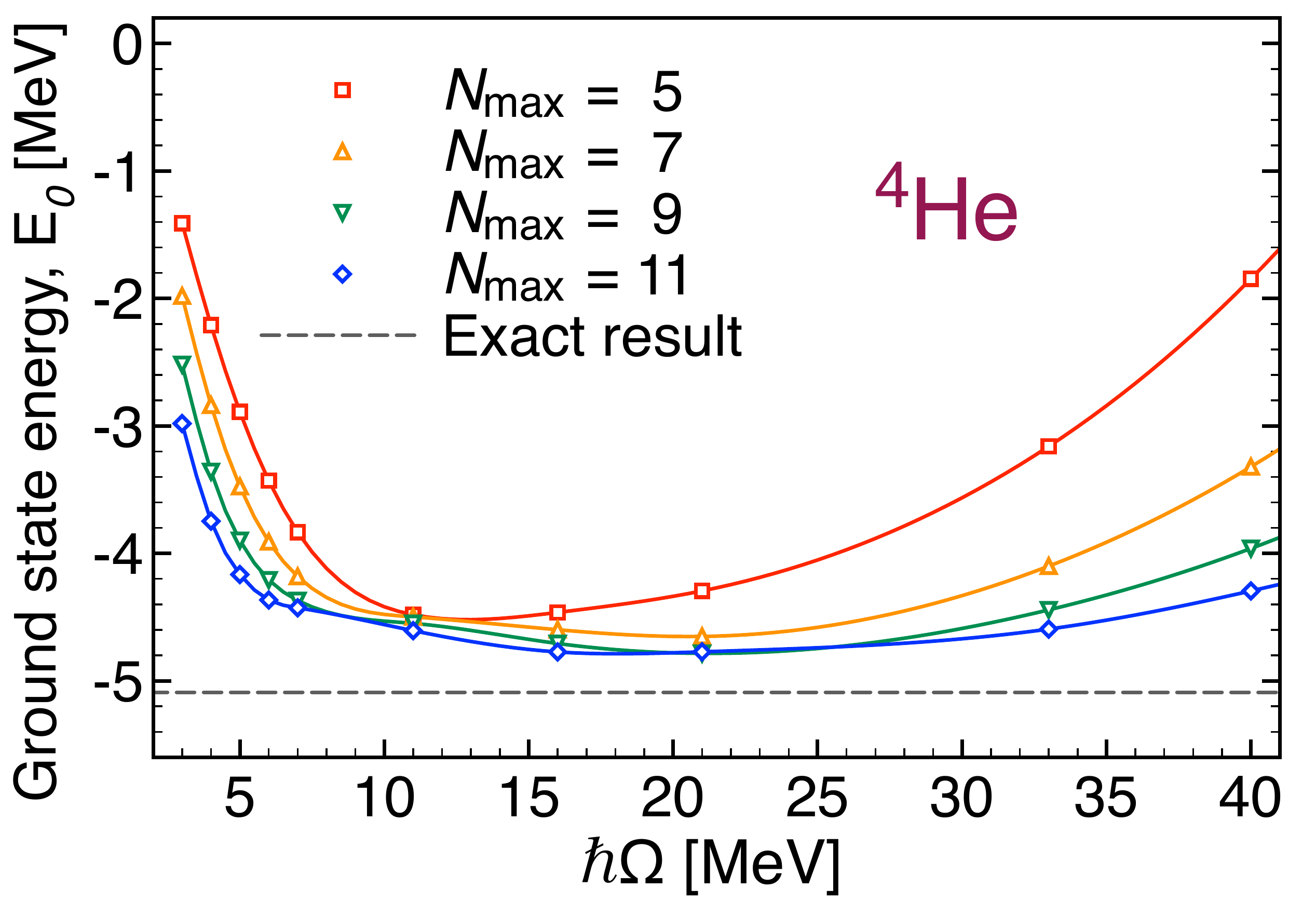} 
\includegraphics[height=0.46\columnwidth,width=0.63\columnwidth,clip=true]{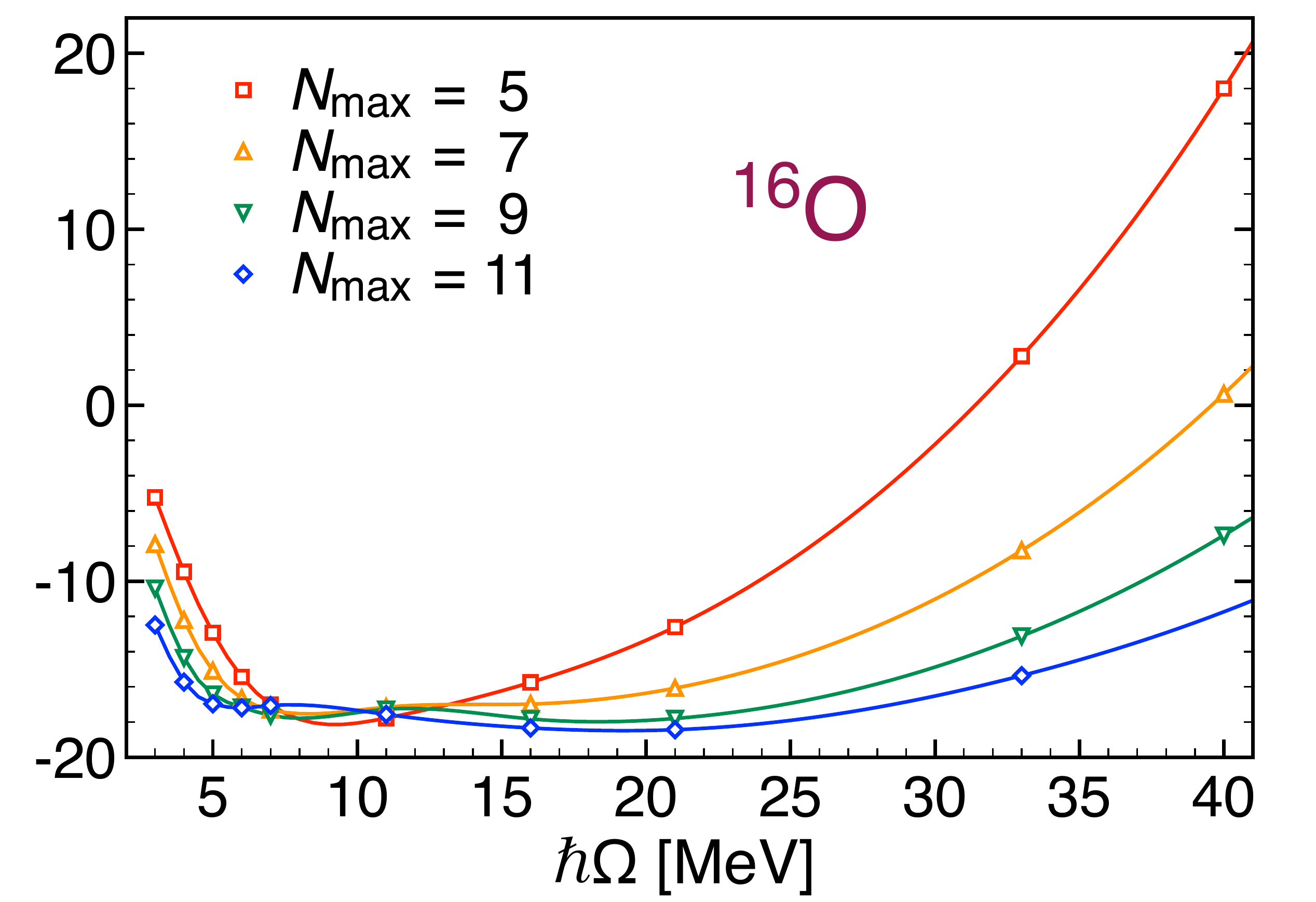} 
\includegraphics[height=0.46\columnwidth,width=0.63\columnwidth,clip=true]{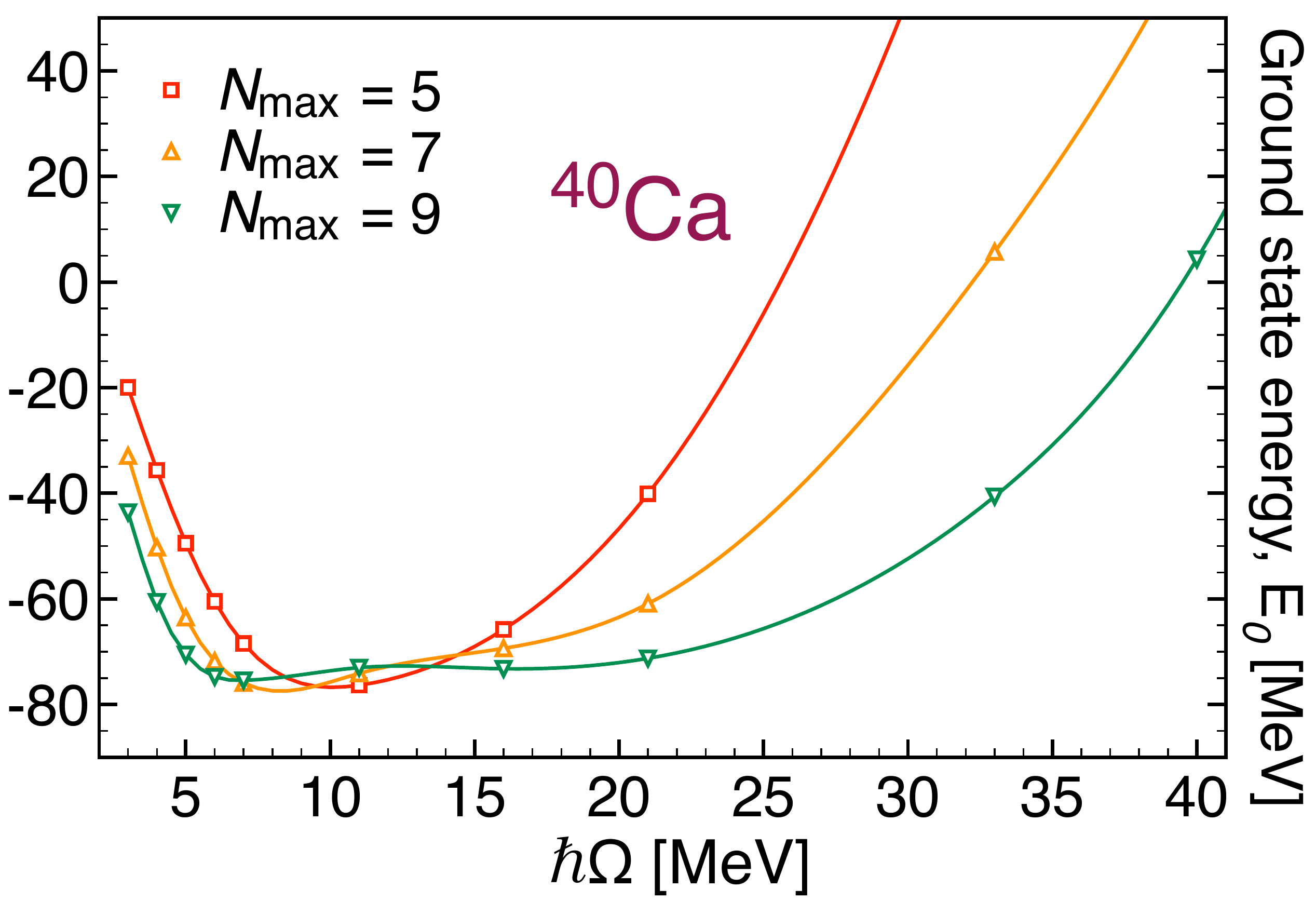}
\caption{(Color online) 
Ground state energy of $^4$He, $^{16}$O and $^{40}$Ca as a function of the harmonic oscillator frequency, $\hbar\Omega$, and the model space size, $\nmax$.
Symbols mark the results  for the HAL469 potential from full self-consistent calculations in the $T^{BGE}(\omega)$ plus ADC(3) approach.
}
\label{fig:He4O16Ca40}
\end{figure*}

Note that the BGE used to generate $T^{BGE}(\omega)$ resums scattering states where at least one nucleon is outside the whole model space. Hence, it does not depend on the isotope being computed, neither it suffers from ambiguities with the choice of the single-particle spectrum at the Fermi surface encountered with the usual G-matrix used in BHF calculations.
Instead, our approach is very similar to the Bloch-Horowitz method of Ref.~\cite{Luu2004BlchHor}, of which our $T^{BGE}(\omega)$ is also a first approximation. The ADC(3) computation accounts for all types of correlations inside the model space, including \emph{all} remaining ladder diagrams. This ensures a complete many-body calculation that accounts for short distance repulsion.

It is well known that short-range repulsion, that is accounted for by Eq.~(\ref{eq:SigMF}), has the double effect of reducing the spectral strength for dominant quasiparticle peaks and of relocating it to large momenta and large quasiparticle energies~\cite{Dickhoff2004ppnp}. 
 Since we cannot currently calculate the location of strength at high momenta, it is not possible to quantify the magnitude of these two effects.
However, they contribute to the Koltun sum rule for total binding energy with opposite signs and must cancel to a large extent. Thus, we chose to neglect both contributions and maintain a static $\Sigma^{(\infty)}$ to solve the Dyson equation. This is currently the major approximation in our calculations and its uncertainty is best estimated from the benchmark on $^4$He below. Resolving this requires a proper extension of the present SCGF formalism and it will be the subject of future work.

{\it Infrared convergence}.
The one-body propagators of $^4$He, $^{16}$O and $^{40}$Ca are calculated in spherical harmonic oscillator spaces of different  frequencies, $\hbar\Omega$, and increasing sizes up to \hbox{$\nmax$=$\max\{2 n + \ell\}$$=$11} (and $\nmax\leq$ 9 for $^{40}$Ca).  
 The scattering matrix $T^{BGE}(\omega)$ is  calculated for each frequency and model space and then used to derive the interactions of Eq.~\eqref{eq:Veff}.
   We subtract the kinetic energy of the center of mass according to Ref.~\cite{Hergert2009Tkin} and calculate the intrinsic ground state energy from $g(\omega)$ using the Koltun sum rule. The  same lattice simulation setup used to generate the HAL469 interaction gives a nucleon mass of $m_N$=1161.1~MeV/c$^2$ in addition to the pseudo-scalar mass of $M_{PS}$=469~MeV/c$^2$. Thus, we employ this value of $m_N$ in all the kinetic energy terms.
Fig.~\ref{fig:He4O16Ca40} displays the ground state energies obtained with our  \hbox{$T^{BGE}(\omega)$} plus ADC(3) method. 
As expected, the complete resummation of ladder diagrams outside the model space tames ultraviolet corrections and results in a rather flat behaviour of the total energies for $\hbar\Omega{}\approx{}$5-20~MeV.   Still, there remain some hints of oscillations with respect to $\hbar\Omega$ that could be linked to the HO truncation and to the neglect of spectral strength at high momenta as explained above.

\begin{figure*}[t]
\includegraphics[height=0.5\columnwidth,clip=true]{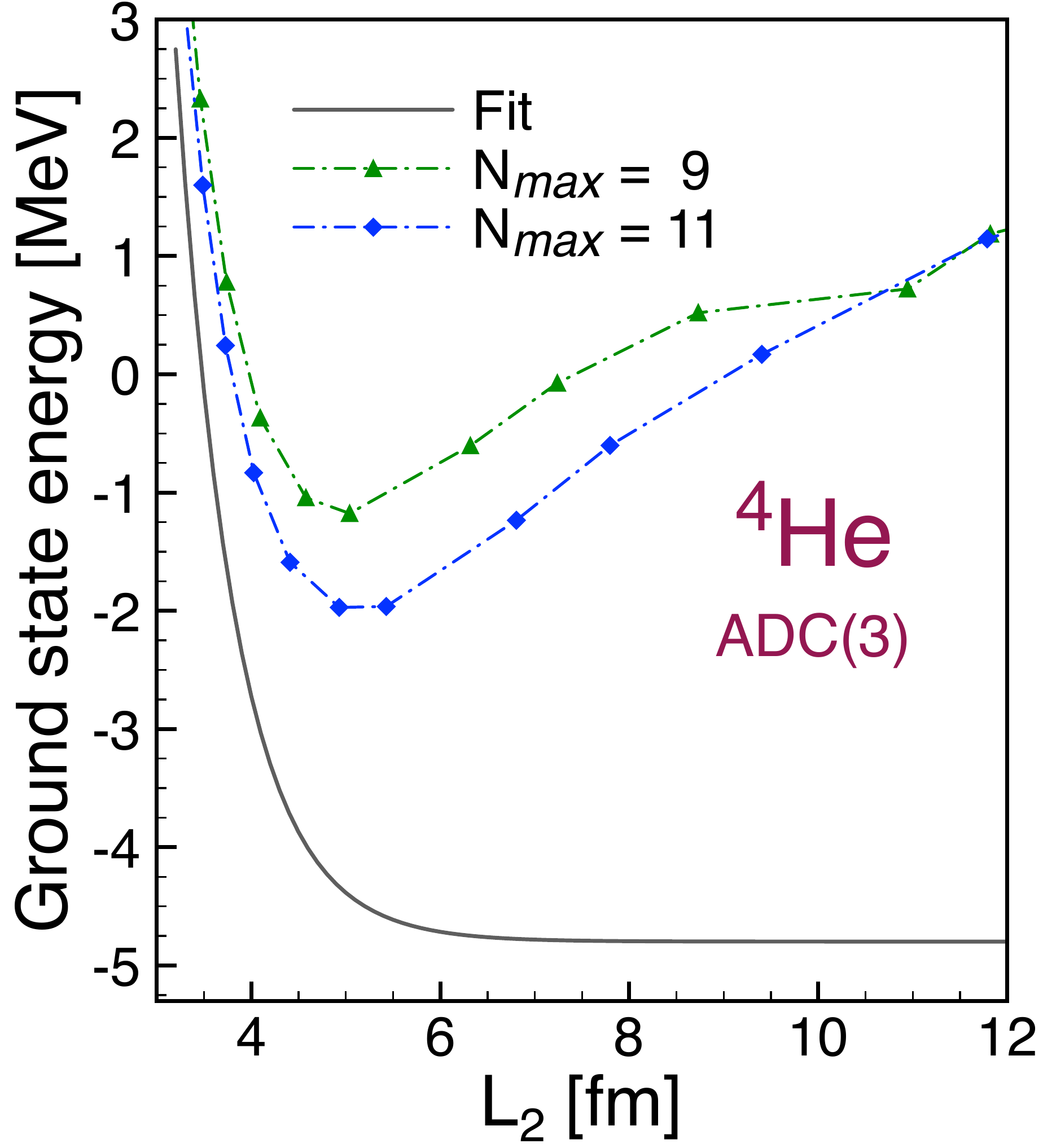} \hspace{.6 cm}
\includegraphics[height=0.5\columnwidth,clip=true]{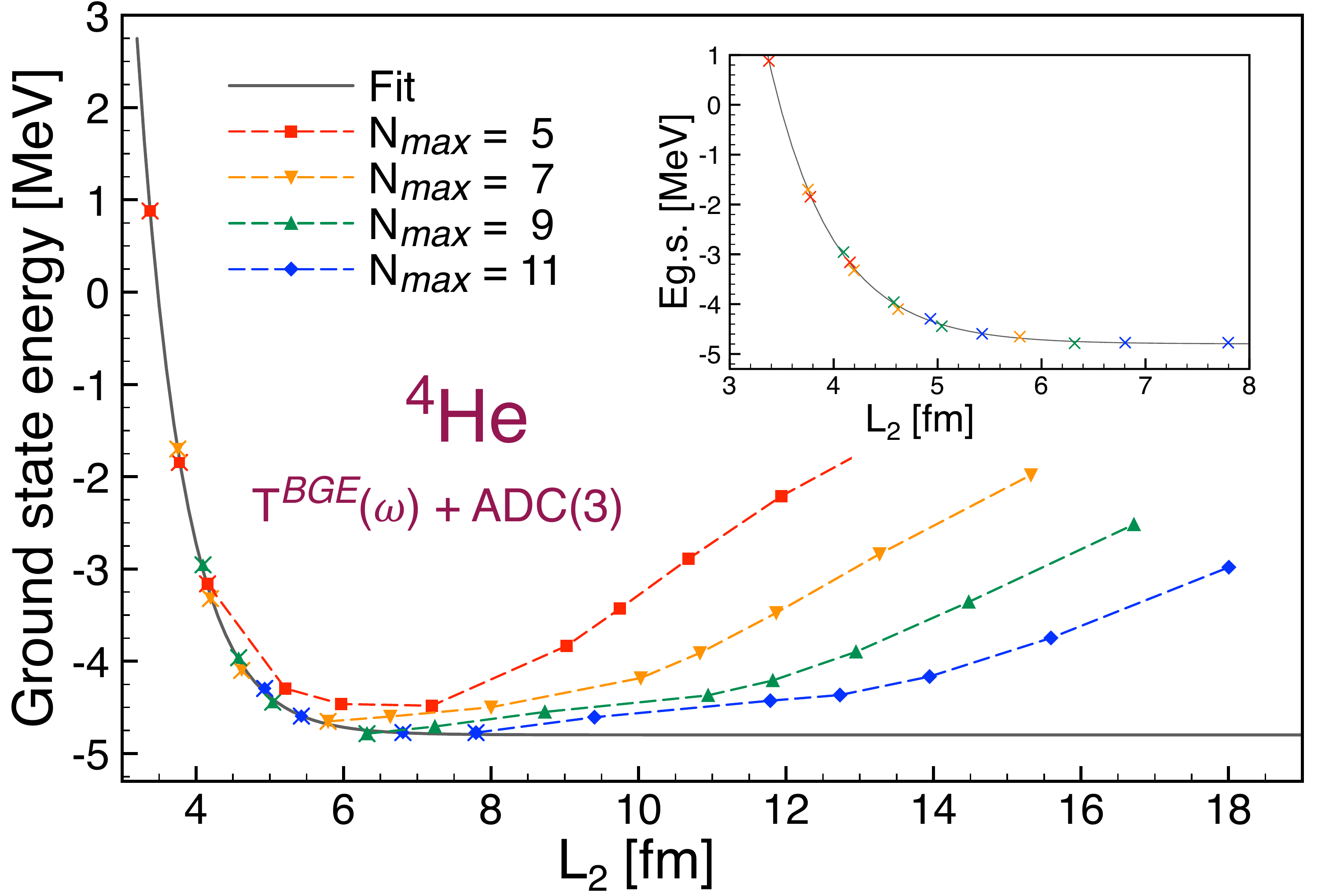} \hspace{.6 cm}
\includegraphics[height=0.5\columnwidth,clip=true]{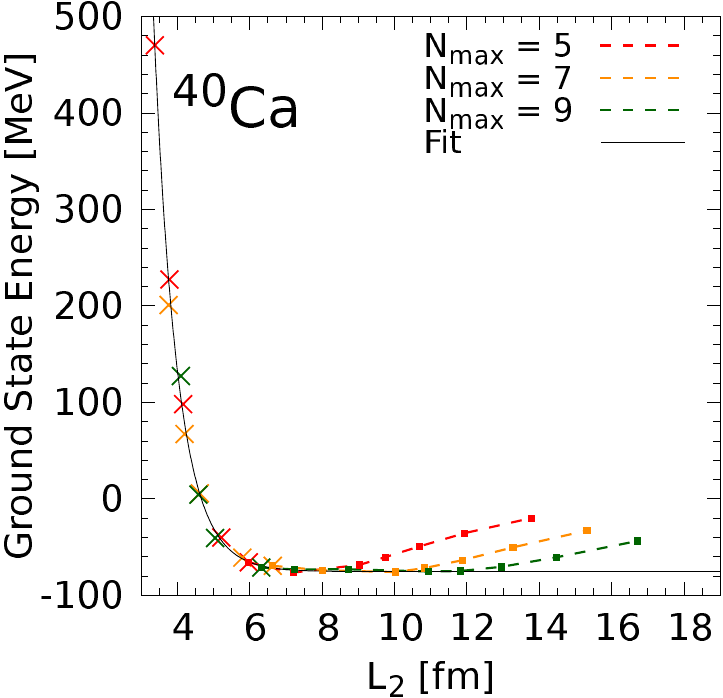}
\caption{(Color online) 
  Calculated ground state energies of $^4$He and $^{40}$Ca for the HAL469 potential as a function of the effective box radius $L_2$. 
 \emph{Left}:~Solution for the bare interaction at $\nmax$=${}$9 and 11 and varying oscillator frequencies without ladders from the excluded space $Q$. 
 \emph{Middle}:~Full calculation, including all ladder diagrams in $Q$. Different colors and broken lines are a guide to the eye
  connecting results of the same  $\nmax$.  The data points included in the fit are marked with crosses and are also show separately in the inset.
 \emph{Right}:~Same as the middle panel but for $^{40}$Ca. 
   For \emph{all panels}, the full black line is the result of the IR extrapolation, with the inclusion of $T^{BGE}$ ladder, according to Eq.~\eqref{eq:IRextr}.}
\label{fig:H4extrap}
\end{figure*}

The effectiveness of the ladder resummation is better recognised by considering the infrared (IR) convergence of the total binding energies, from where one can also extrapolate to a complete set. 
Ref.~\cite{More2013InfraredExtr} established that a harmonic oscillator model space, of frequency $\hbar\Omega$ and truncated to the first $\nmax+1$ shells, behaves as a hard wall spherical box of radius
\begin{align}
  L_2 ={}& \sqrt{2 ( \nmax + 3/2 + 2)} \, b \; ,
  \label{eq:L2}
\end{align}
where $b\equiv\sqrt{\hbar c^2/ m_N \Omega}$ is the oscillator length.  Given a soft interaction that is independent of the model space, if the frequency is large enough (i.e., $b$  is small) then ultraviolet (UV) degrees of freedom are converged. In this case, the calculated ground state energies are expected to converge exponentially when increasing the effective radius $L_2$:
\begin{align}
  E_0^A[\nmax,\hbar\Omega] ={}& E_{\infty} ~+ ~C \; e^{- 2 \, k_\infty \,  L_2}.
\label{eq:IRextr}
\end{align}
For the bare HAL469 interaction, if we use the SCGF without ladder sums outside the model space, the extrapolation according to Eq.~\eqref{eq:IRextr} will fail because the short-distance repulsion requires extremely large $\nmax$ ($\gg$ 20) to reach UV convergence, while our many-body space $P$ is limited to $\nmax$=11. 
This is shown by the left panel of Fig.~\ref{fig:H4extrap}.
However, our complete calculations  use the BGE to resum all missing ladder diagrams within $Q$. Adding these to the two-particle ladders that are generated by the ADC(3) leads to a \emph{complete} diagonalization of short-distance degrees of freedom, independently of the choice of $P$.   
The resulting dependence of the ground state energy on $L_2$ is shown by the middle panel and it now follows the behaviour dictated by Eq.~\eqref{eq:IRextr}.
Note that single particle energies are still needed to calculate $T^{BGE}(\omega)$ but these can be identified with the free particle spectrum in the space $Q$, at large momenta.
Accordingly, the flat region in Fig.~\ref{fig:H4extrap} becomes broader as we increase $\nmax$ (that is, when the boundary between the $P$ and $Q$ spaces moves away form the Fermi surface).

The binding energy of $^4$He for  HAL469 was found to be \hbox{-5.09~MeV} with the \emph{exact} Stochastic Variational calculations~\cite{Nemura2016WSPC_procs}, which we will use to benchmark our approach.  
The solid lines in Fig.~\ref{fig:H4extrap} are the result of a nonlinear least-squares fit to Eq.~\eqref{eq:IRextr}. The points diverging from the exponential behaviour  at large $L_2$ are assumed not to be UV converged due to the above-mentioned approximations and are excluded from the fit but are still shown in the figure.
From calculations up to $\hbar\Omega$=50~MeV and the IR extrapolation, we estimate a converged binding energy of 4.80(3) MeV for $^4$He, where the error corresponds to the uncertainties in the extrapolation.
The calculations for the other isotopes converge similarly  to $^4$He, and we show the IR extrapolation of $^{40}$Ca in the right panel of Fig.~\ref{fig:H4extrap}, for completeness. The figure  is also indicative that this methodology can be successfully applied to heavier nuclei.
These results clearly show that SCGF can handle relatively hard potentials by resumming of ladders through the  Bethe-Goldstone equation
and they give confidence that the short-range repulsion of HAL469 is accounted for accurately.

{\it Results}.
The calculated ground state energies of $^4$He, $^{16}$O and $^{40}$Ca are summarised in Tab.~\ref{tab:He4O16Ca40},  together with  BHF results obtained with the same gap choice and methods of Ref.~\cite{Inoue:2014ipa}.  
For $^4$He, the complete $T^{BGE}(\omega)$+ADC(3) result deviates from the exact solution for  $^4$He by less than~10\%.  Since the SCGF approach resums linked diagrams, and thus it is size extensive, one should expect that similar errors will apply for larger isotopes~\cite{shavitt2009many}.
Thus, Tab.~\ref{tab:He4O16Ca40} shows both the uncertainties in the IR extrapolation~\cite{More2013InfraredExtr} and an error for the many-body truncations, for we which we take a conservative estimate of 10\% based on our finding for $^4$He.
The SCGF results are sensibly less bound than our previous BHF results~\cite{Inoue:2014ipa} and we interpret this as a limitation of BHF theory. 

%

\begin{table}[b]
\begin{center}
\begin{tabular}{lcccccc}
\hline
\hline
$E^A_0$ [MeV] &&  $^4$He  && $^{16}$O  && $^{40}$Ca  \\
\hline
BHF~\cite{Inoue:2014ipa} &&  -8.2  && -34.7   &&  -112.7  \\
$T^{BGE}(\omega)$+ADC(3) &&   -4.80(0.03)  && -17.9 (0.3) (1.8)  &&  -75.4 (6.7) (7.5)  \\
Exact calc.~\cite{Nemura2016WSPC_procs} && -5.09   && --  && --  \\
Experiment  &&  -28.3   &&  -127.7  &&   -342.0 \\
\hline
\multicolumn{3}{l}{Separation into $^4$He clusters: }   && -2.46 (0.3) (1.8)  && 24.5 (6.7) (7.5)  \\
\hline
\hline
\end{tabular}
\caption{
Ground state energies of  $^4$He, $^{16}$O and $^{40}$Ca at $M_{PS}$=469~MeV/c$^2$ obtained from the HAL469 interaction.  The `$T^{BGE}(\omega)$+ADC(3)' results of the present work are compared to BHF and to the exact solution.
The last line is the breakup energy for splitting the system in  $^4$He clusters (of total energy $A/4\times$5.09~MeV). }
\label{tab:He4O16Ca40}
\end{center} 
\end{table}

%


A key feature of our calculations is the use of an harmonic oscillator space, which effectively confines all nucleons.  The last line of Tab.~\ref{tab:He4O16Ca40} reports the deduced breakup energies for separating the computed ground states into infinitely distant $^4$He clusters.
The $^{16}$O is unstable with respect to 4-$\alpha$ break up, by~$\approx{}$2.5~MeV. Allowing an error in our  binding energies of more than 10\% could make oxygen bound but only very weakly. This is in contrast to the experimental results, at the physical quarks masses, where the 4-$\alpha$ breakup requires 14.4~MeV.   On the other hand, $^{40}$Ca is stable with respect to breakup in $\alpha$ particles by $\approx{}$24~MeV.  
We expect that these observations are rather robust even when we consider the (LQCD) statistical errors in the HAL469 interaction.
While such statistical fluctuations introduce additional $\sim$10\% errors on binding energies~\cite{Inoue:2014ipa},
they are expected to be strongly correlated among $^4$He, $^{16}$O and $^{40}$Ca.
Hence, for QCD in the SU(3) limit at $M_{PS}$=469~MeV/c$^2$, we find that the deuteron is unbound~\cite{Inoue:2011ai} and $^{16}$O is only just slightly above the threshold for $\alpha$ breakup, while  $^4$He and $^{40}$Ca are instead bound. The HAL469 interaction has the lowest $M_{PS}$ value among those considered in Refs.~\cite{Inoue:2010es,Inoue:2011ai}, while from Ref.~\cite{Inoue:2013nfe} we know that it is the only one saturating nuclear matter (although not at the physical saturation point). Moreover, we have tested that SCGF attempts at calculating asymmetric isotopes, like $^{28}$O, predict strongly unbound systems even for HAL469. All these results together suggest that, when lowering of the pion mass toward its physical value,  closed shell isotopes are created at first around the traditional magic numbers. 
This hypothesis should also be seen in the light of the  limitations in the present HAL469 Hamiltonian, which was built to include only the $^1$S$_0$, $^3$S$_1$ and $^3$D$_1$ partial waves and therefore neglects the three-body forces and spin-orbit interactions. The missing $P$ waves and Coulomb force are repulsive but could be compensated by an attracting three-body force.

\begin{figure}[t]
\includegraphics[height=0.46\columnwidth,clip=true]{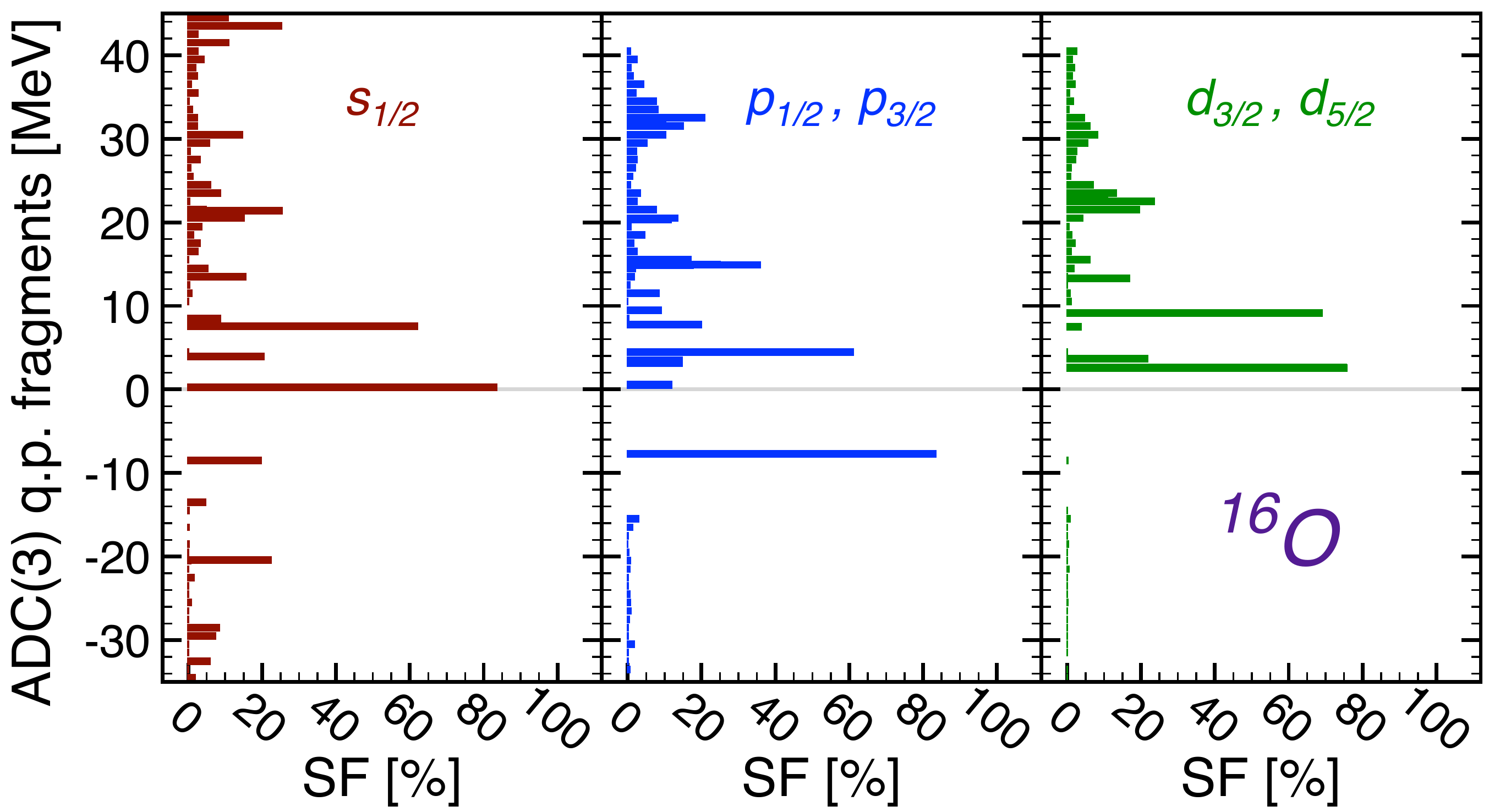}
\caption{(Color online) 
 Single particle spectral strength distribution of $^{16}$O obtained from the dressed propagator  in the full \hbox{$T^{BGE}(\omega)$} plus ADC(3) approach. Each panel displays  partial waves of different angular momenta. The vertical axes give the quasiparticle energies (that is, the poles of Eq.~\eqref{eq:gLeh}), while the length of the horizontal bars give the calculated spectroscopic factors.
}
\label{fig:O16SF}
\end{figure}

\begin{table}[b]
\begin{center}
\begin{tabular}{clccccccccc}
\hline
\hline
~~&&~& &~~~~& $^{4}$He &~~~~& $^{16}$O &~~~~& $^{40}$Ca & ~~\\
\hline
&$r_{pt-matter}$[fm]:&&
BHF~\cite{Inoue:2014ipa} &\qquad& 2.09  &\qquad& 2.35  &\qquad&  2.78  \\
&&& HF && 1.62 && 2.39  && 2.78  \\
&&&$T^{BGE}(\omega)$ + ADC(3) && 1.67 && 2.64  && 2.97  \\
\hline
&$r_{charge}$[fm]: && $T^{BGE}(\omega)$ + ADC(3) && 1.89  && 2.79  && 3.10  \\
&&& Experiment~\cite{DEVRIES1987chradii,Angeli2013chradii} && 1.67 && 2.73  && 3.48  \\
\hline
\hline
\end{tabular}
\end{center} 
\caption{Computed matter and charge radii of $^{16}$O and $^{40}$Ca  using M$_{PS}$=469~MeV for $\nmax$=11 and $\hbar\Omega{}$=${}$11~MeV. Results are given for
different levels of approximations and the charge radii from the full $T^{BGE}(\omega)$ plus ADC(3) are compared to the experimental values.
For charge radii, we assumed the physical charge distributions of the proton and the neutron (see Ref.~\cite{Cipollone2015OxSpFnct} for details).
}
\label{tab:radii}
\end{table} 

Figure~\ref{fig:O16SF} demonstrates the spectral strength distribution of $^{16}$O  obtained
for $\nmax$=11 and $\hbar\Omega{}$=${}$11~MeV.  Quasiparticle fragments corresponding to spin-orbit partners do not split due to the absence of a spin-orbit term in HAL469. Otherwise, all the remaining qualitative features of the experimental spectral distribution are seen also for the $M_{PS}$=469~MeV/c$^2$.
The root mean square radii are given Tab.~\ref{tab:radii} for the same model space and oscillator frequency. 
 Although the total binding energies are 15-20\% of the experimental value (Tab.~\ref{tab:He4O16Ca40}),
 the computed charge radii are about the same as the experiment. 
   This is due to the fact that
 the heavy nucleon mass ($m_N$=1161.1 MeV/${\rm c}^2$) used here 
reduces the
  motion of the nucleons inside the nuclei and counterbalances the effect of weak attraction 
of the HAL469 potential.
%
%
 We have also checked the dependence of the computed radii on the effective model space size, Eq.~\eqref{eq:L2}, and found a rather flat converged region for all three isotopes, although the values still oscillate by about 0.05~fm with changing $L_2$. The values in Tab.~\ref{tab:radii} are all calculated in the middle of this plateau. 
The HF approach of Eq.~\eqref{eq:GHF} and the standard BHF give similar radii in spite that they predict very different binding energies. The final radii are then increased by many-body correlations and, for all nuclei, the full $T^{BGE}(\omega)$ plus ADC(3) calculations pushes the matter distribution to larger discances. However, we note that accounting for the neglected high-momentum components---as discussed below Eq.~\eqref{eq:Veff}---tends to enhance the central density and would slightly reduce the calculated radii~\cite{MuthSickPRC2004_rohr}.

%

%

{\it Summary}.
We investigated the use of the BGE in an {\em ab~initio} approach and used it to resum missing two-nucleon scattering diagrams outside the  usual  truncations of the many-body space, while the full ADC(3) method has been retained within the model space itself.
A benchmark on $^4$He shows that the present implementation  works relatively well and it 
allows  to  solve the self-consistent Green's function for the HAL~QCD potentials derived from Lattice QCD.
An investigation of the IR convergence of the ground state energies, following the work of Ref.~\cite{More2013InfraredExtr}, indicates that SCGF can handle relatively hard potentials such as the HAL469, even masses as large as A=40.
This opens a new path that allows full {\em ab initio} calculations of  large nuclei even with hard nuclear interactions.

The present accuracy is sufficient to make quantitative statements on doubly magic nuclei, which are less bound compared to earlier BHF estimates for the HAL QCD potentials. Here, we have found that the behaviour when lowering the pion mass towards its physical value is consistent with the idea that nuclei near to the traditional magic numbers are formed at first.  At $M_{PS}$=469~MeV/c$^2$, in the SU(3) limit of QCD,  both $^4$He and $^{40}$Ca have bound ground states while the deuteron is unbound and $^{16}$O  is likely to decay into four separate alpha particles. However, $^{16}$O is already close to become bound.
This suggests that the region of $M_{PS}\sim{}$500~MeV/c$^2$ marks a transition between an unbound nuclear chart and the emergence of bound isotopes.  Further studies at lower pion masses will be pivotal to test these findings and should be possible in the near future since LQCD simulations for nuclear  and hyperon forces down to physical quark masses are currently underway~\cite{Doi:2015oha,Ishii:2016zsf,Sasaki:2016gpc}. 

%

%
Important future work will also be the inclusion the spin-orbit as well as three-nucleon forces.  Proof of principle LQCD calculations for these interactions are available~\cite{Murano:2013xxa,Doi:2011gq} and follow the hierarchy of nuclear forces with three-nucleon terms smaller than the NN contributions.

{\it Acknowledgements}. 
We thank the HAL QCD Collaboration for providing the HAL469$_{\hbox{\small  \em SU(3)}}$ interaction.  
Computations of the \hbox{$T^{BGE}(\omega)$} were performed using the CENS codes from Ref.~\cite{engeland2008cens}.
This work was supported in part by the United Kingdom Science and Technology Facilities Council (STFC) under Grants No. ST/L005743/1 and ST/L005816/1, by the Royal Society International Exchanges Grant No.~IE150305,
by Japanese Grant-in-Aid for Scientific Research (JP15K17667and (C)26400281),
by MEXT Strategic Program for Innovative Research (SPIRE) Field 5,  
by a priority issue (Elucidation of the fundamental laws and evolution of the universe) to be tackled by using Post ``K" Computer, 
and by the  Joint Institute for Computational Fundamental Science (JICFuS).
T.D and T.H. were supported partially by the RIKEN iTHEMS Program and iTHES Project.
Calculations were performed at the DiRAC Complexity system at the University of Leicester (BIS National E- infrastructure capital Grant No. ST/K000373/1 and STFC Grant No. ST/K0003259/1).


\bibliography{HALGF_Refs}

\begin{thebibliography}{62}%
\makeatletter
\providecommand \@ifxundefined [1]{%
 \@ifx{#1\undefined}
}%
\providecommand \@ifnum [1]{%
 \ifnum #1\expandafter \@firstoftwo
 \else \expandafter \@secondoftwo
 \fi
}%
\providecommand \@ifx [1]{%
 \ifx #1\expandafter \@firstoftwo
 \else \expandafter \@secondoftwo
 \fi
}%
\providecommand \natexlab [1]{#1}%
\providecommand \enquote  [1]{``#1''}%
\providecommand \bibnamefont  [1]{#1}%
\providecommand \bibfnamefont [1]{#1}%
\providecommand \citenamefont [1]{#1}%
\providecommand \href@noop [0]{\@secondoftwo}%
\providecommand \href [0]{\begingroup \@sanitize@url \@href}%
\providecommand \@href[1]{\@@startlink{#1}\@@href}%
\providecommand \@@href[1]{\endgroup#1\@@endlink}%
\providecommand \@sanitize@url [0]{\catcode `\\12\catcode `\$12\catcode
  `\&12\catcode `\#12\catcode `\^12\catcode `\_12\catcode `\%12\relax}%
\providecommand \@@startlink[1]{}%
\providecommand \@@endlink[0]{}%
\providecommand \url  [0]{\begingroup\@sanitize@url \@url }%
\providecommand \@url [1]{\endgroup\@href {#1}{\urlprefix }}%
\providecommand \urlprefix  [0]{URL }%
\providecommand \Eprint [0]{\href }%
\providecommand \doibase [0]{http://dx.doi.org/}%
\providecommand \selectlanguage [0]{\@gobble}%
\providecommand \bibinfo  [0]{\@secondoftwo}%
\providecommand \bibfield  [0]{\@secondoftwo}%
\providecommand \translation [1]{[#1]}%
\providecommand \BibitemOpen [0]{}%
\providecommand \bibitemStop [0]{}%
\providecommand \bibitemNoStop [0]{.\EOS\space}%
\providecommand \EOS [0]{\spacefactor3000\relax}%
\providecommand \BibitemShut  [1]{\csname bibitem#1\endcsname}%
\let\auto@bib@innerbib\@empty
\bibitem [{\citenamefont {Borsanyi}\ \emph {et~al.}(2015)\citenamefont
  {Borsanyi}, \citenamefont {Durr}, \citenamefont {Fodor}, \citenamefont
  {Hoelbling}, \citenamefont {Katz}, \citenamefont {Krieg}, \citenamefont
  {Lellouch}, \citenamefont {Lippert}, \citenamefont {Portelli}, \citenamefont
  {Szabo},\ and\ \citenamefont {Toth}}]{Borsanyi:2014jba}%
  \BibitemOpen
  \bibfield  {author} {\bibinfo {author} {\bibfnamefont {S.}~\bibnamefont
  {Borsanyi}}, \bibinfo {author} {\bibfnamefont {S.}~\bibnamefont {Durr}},
  \bibinfo {author} {\bibfnamefont {Z.}~\bibnamefont {Fodor}}, \bibinfo
  {author} {\bibfnamefont {C.}~\bibnamefont {Hoelbling}}, \bibinfo {author}
  {\bibfnamefont {S.~D.}\ \bibnamefont {Katz}}, \bibinfo {author}
  {\bibfnamefont {S.}~\bibnamefont {Krieg}}, \bibinfo {author} {\bibfnamefont
  {L.}~\bibnamefont {Lellouch}}, \bibinfo {author} {\bibfnamefont
  {T.}~\bibnamefont {Lippert}}, \bibinfo {author} {\bibfnamefont
  {A.}~\bibnamefont {Portelli}}, \bibinfo {author} {\bibfnamefont {K.~K.}\
  \bibnamefont {Szabo}}, \ and\ \bibinfo {author} {\bibfnamefont {B.~C.}\
  \bibnamefont {Toth}},\ }\href {\doibase 10.1126/science.1257050} {\bibfield
  {journal} {\bibinfo  {journal} {Science}\ }\textbf {\bibinfo {volume}
  {347}},\ \bibinfo {pages} {1452} (\bibinfo {year} {2015})}\BibitemShut
  {NoStop}%
\bibitem [{\citenamefont {Beane}\ \emph {et~al.}(2012)\citenamefont {Beane},
  \citenamefont {Chang}, \citenamefont {Detmold}, \citenamefont {Lin},
  \citenamefont {Luu}, \citenamefont {Orginos}, \citenamefont {Parre\~no},
  \citenamefont {Savage}, \citenamefont {Torok},\ and\ \citenamefont
  {Walker-Loud}}]{Beane:2011iw}%
  \BibitemOpen
  \bibfield  {author} {\bibinfo {author} {\bibfnamefont {S.~R.}\ \bibnamefont
  {Beane}}, \bibinfo {author} {\bibfnamefont {E.}~\bibnamefont {Chang}},
  \bibinfo {author} {\bibfnamefont {W.}~\bibnamefont {Detmold}}, \bibinfo
  {author} {\bibfnamefont {H.~W.}\ \bibnamefont {Lin}}, \bibinfo {author}
  {\bibfnamefont {T.~C.}\ \bibnamefont {Luu}}, \bibinfo {author} {\bibfnamefont
  {K.}~\bibnamefont {Orginos}}, \bibinfo {author} {\bibfnamefont
  {A.}~\bibnamefont {Parre\~no}}, \bibinfo {author} {\bibfnamefont {M.~J.}\
  \bibnamefont {Savage}}, \bibinfo {author} {\bibfnamefont {A.}~\bibnamefont
  {Torok}}, \ and\ \bibinfo {author} {\bibfnamefont {A.}~\bibnamefont
  {Walker-Loud}} (\bibinfo {collaboration} {NPLQCD Collaboration}),\ }\href
  {\doibase 10.1103/PhysRevD.85.054511} {\bibfield  {journal} {\bibinfo
  {journal} {Phys. Rev. D}\ }\textbf {\bibinfo {volume} {85}},\ \bibinfo
  {pages} {054511} (\bibinfo {year} {2012})}\BibitemShut {NoStop}%
\bibitem [{\citenamefont {Yamazaki}\ \emph {et~al.}(2012)\citenamefont
  {Yamazaki}, \citenamefont {Ishikawa}, \citenamefont {Kuramashi},\ and\
  \citenamefont {Ukawa}}]{Yamazaki:2012hi}%
  \BibitemOpen
  \bibfield  {author} {\bibinfo {author} {\bibfnamefont {T.}~\bibnamefont
  {Yamazaki}}, \bibinfo {author} {\bibfnamefont {K.-i.}\ \bibnamefont
  {Ishikawa}}, \bibinfo {author} {\bibfnamefont {Y.}~\bibnamefont {Kuramashi}},
  \ and\ \bibinfo {author} {\bibfnamefont {A.}~\bibnamefont {Ukawa}},\ }\href
  {\doibase 10.1103/PhysRevD.86.074514} {\bibfield  {journal} {\bibinfo
  {journal} {Phys. Rev. D}\ }\textbf {\bibinfo {volume} {86}},\ \bibinfo
  {pages} {074514} (\bibinfo {year} {2012})}\BibitemShut {NoStop}%
\bibitem [{\citenamefont {Yamazaki}\ \emph {et~al.}(2015)\citenamefont
  {Yamazaki}, \citenamefont {Ishikawa}, \citenamefont {Kuramashi},\ and\
  \citenamefont {Ukawa}}]{Yamazaki:2015asa}%
  \BibitemOpen
  \bibfield  {author} {\bibinfo {author} {\bibfnamefont {T.}~\bibnamefont
  {Yamazaki}}, \bibinfo {author} {\bibfnamefont {K.-i.}\ \bibnamefont
  {Ishikawa}}, \bibinfo {author} {\bibfnamefont {Y.}~\bibnamefont {Kuramashi}},
  \ and\ \bibinfo {author} {\bibfnamefont {A.}~\bibnamefont {Ukawa}},\ }\href
  {\doibase 10.1103/PhysRevD.92.014501} {\bibfield  {journal} {\bibinfo
  {journal} {Phys. Rev. D}\ }\textbf {\bibinfo {volume} {92}},\ \bibinfo
  {pages} {014501} (\bibinfo {year} {2015})}\BibitemShut {NoStop}%
\bibitem [{\citenamefont {Orginos}\ \emph {et~al.}(2015)\citenamefont
  {Orginos}, \citenamefont {Parre\~no}, \citenamefont {Savage}, \citenamefont
  {Beane}, \citenamefont {Chang},\ and\ \citenamefont
  {Detmold}}]{Orginos:2015aya}%
  \BibitemOpen
  \bibfield  {author} {\bibinfo {author} {\bibfnamefont {K.}~\bibnamefont
  {Orginos}}, \bibinfo {author} {\bibfnamefont {A.}~\bibnamefont {Parre\~no}},
  \bibinfo {author} {\bibfnamefont {M.~J.}\ \bibnamefont {Savage}}, \bibinfo
  {author} {\bibfnamefont {S.~R.}\ \bibnamefont {Beane}}, \bibinfo {author}
  {\bibfnamefont {E.}~\bibnamefont {Chang}}, \ and\ \bibinfo {author}
  {\bibfnamefont {W.}~\bibnamefont {Detmold}} (\bibinfo {collaboration} {NPLQCD
  Collaboration}),\ }\href {\doibase 10.1103/PhysRevD.92.114512} {\bibfield
  {journal} {\bibinfo  {journal} {Phys. Rev. D}\ }\textbf {\bibinfo {volume}
  {92}},\ \bibinfo {pages} {114512} (\bibinfo {year} {2015})}\BibitemShut
  {NoStop}%
\bibitem [{\citenamefont {Berkowitz}\ \emph {et~al.}(2015)\citenamefont
  {Berkowitz}, \citenamefont {Kurth}, \citenamefont {Nicholson}, \citenamefont
  {Joo}, \citenamefont {Rinaldi}, \citenamefont {Strother}, \citenamefont
  {Vranas},\ and\ \citenamefont {Walker-Loud}}]{Berkowitz:2015eaa}%
  \BibitemOpen
  \bibfield  {author} {\bibinfo {author} {\bibfnamefont {E.}~\bibnamefont
  {Berkowitz}}, \bibinfo {author} {\bibfnamefont {T.}~\bibnamefont {Kurth}},
  \bibinfo {author} {\bibfnamefont {A.}~\bibnamefont {Nicholson}}, \bibinfo
  {author} {\bibfnamefont {B.}~\bibnamefont {Joo}}, \bibinfo {author}
  {\bibfnamefont {E.}~\bibnamefont {Rinaldi}}, \bibinfo {author} {\bibfnamefont
  {M.}~\bibnamefont {Strother}}, \bibinfo {author} {\bibfnamefont {P.~M.}\
  \bibnamefont {Vranas}}, \ and\ \bibinfo {author} {\bibfnamefont
  {A.}~\bibnamefont {Walker-Loud}},\ }\href@noop {} {\  (\bibinfo {year}
  {2015})},\ \Eprint {http://arxiv.org/abs/1508.00886} {arXiv:1508.00886
  [hep-lat]} \BibitemShut {NoStop}%
\bibitem [{\citenamefont {Iritani}\ \emph {et~al.}(2016)\citenamefont
  {Iritani}, \citenamefont {Doi}, \citenamefont {Aoki}, \citenamefont {Gongyo},
  \citenamefont {Hatsuda}, \citenamefont {Ikeda}, \citenamefont {Inoue},
  \citenamefont {Ishii}, \citenamefont {Murano}, \citenamefont {Nemura},\ and\
  \citenamefont {Sasaki}}]{Iritani:2016jie}%
  \BibitemOpen
  \bibfield  {author} {\bibinfo {author} {\bibfnamefont {T.}~\bibnamefont
  {Iritani}}, \bibinfo {author} {\bibfnamefont {T.}~\bibnamefont {Doi}},
  \bibinfo {author} {\bibfnamefont {S.}~\bibnamefont {Aoki}}, \bibinfo {author}
  {\bibfnamefont {S.}~\bibnamefont {Gongyo}}, \bibinfo {author} {\bibfnamefont
  {T.}~\bibnamefont {Hatsuda}}, \bibinfo {author} {\bibfnamefont
  {Y.}~\bibnamefont {Ikeda}}, \bibinfo {author} {\bibfnamefont
  {T.}~\bibnamefont {Inoue}}, \bibinfo {author} {\bibfnamefont
  {N.}~\bibnamefont {Ishii}}, \bibinfo {author} {\bibfnamefont
  {K.}~\bibnamefont {Murano}}, \bibinfo {author} {\bibfnamefont
  {H.}~\bibnamefont {Nemura}}, \ and\ \bibinfo {author} {\bibfnamefont
  {K.}~\bibnamefont {Sasaki}},\ }\href {\doibase 10.1007/JHEP10(2016)101}
  {\bibfield  {journal} {\bibinfo  {journal} {Journal of High Energy Physics}\
  }\textbf {\bibinfo {volume} {2016}},\ \bibinfo {pages} {101} (\bibinfo {year}
  {2016})}\BibitemShut {NoStop}%
\bibitem [{\citenamefont {Aoki}\ \emph {et~al.}(2016)\citenamefont {Aoki},
  \citenamefont {Doi},\ and\ \citenamefont {Iritani}}]{Aoki:2016dmo}%
  \BibitemOpen
  \bibfield  {author} {\bibinfo {author} {\bibfnamefont {S.}~\bibnamefont
  {Aoki}}, \bibinfo {author} {\bibfnamefont {T.}~\bibnamefont {Doi}}, \ and\
  \bibinfo {author} {\bibfnamefont {T.}~\bibnamefont {Iritani}},\ }\bibfield
  {booktitle} {\emph {\bibinfo {booktitle} {{Proceedings, 34th International
  Symposium on Lattice Field Theory (Lattice 2016): Southampton, UK, July
  24-30, 2016}}},\ }\href@noop {} {\bibfield  {journal} {\bibinfo  {journal}
  {PoS LATTICE}\ }\textbf {\bibinfo {volume} {2016}},\ \bibinfo {pages} {109}
  (\bibinfo {year} {2016})},\ \Eprint {http://arxiv.org/abs/1610.09763}
  {arXiv:1610.09763 [hep-lat]} \BibitemShut {NoStop}%
\bibitem [{\citenamefont {Iritani}(2016)}]{Iritani:2016xmx}%
  \BibitemOpen
  \bibfield  {author} {\bibinfo {author} {\bibfnamefont {T.}~\bibnamefont
  {Iritani}},\ }\bibfield  {booktitle} {\emph {\bibinfo {booktitle}
  {{Proceedings, 34th International Symposium on Lattice Field Theory (Lattice
  2016): Southampton, UK, July 24-30, 2016}}},\ }\href@noop {} {\bibfield
  {journal} {\bibinfo  {journal} {PoS LATTICE}\ }\textbf {\bibinfo {volume}
  {2016}},\ \bibinfo {pages} {107} (\bibinfo {year} {2016})},\ \Eprint
  {http://arxiv.org/abs/1610.09779} {arXiv:1610.09779 [hep-lat]} \BibitemShut
  {NoStop}%
\bibitem [{\citenamefont {Ishii}\ \emph {et~al.}(2007)\citenamefont {Ishii},
  \citenamefont {Aoki},\ and\ \citenamefont {Hatsuda}}]{Ishii:2006ec}%
  \BibitemOpen
  \bibfield  {author} {\bibinfo {author} {\bibfnamefont {N.}~\bibnamefont
  {Ishii}}, \bibinfo {author} {\bibfnamefont {S.}~\bibnamefont {Aoki}}, \ and\
  \bibinfo {author} {\bibfnamefont {T.}~\bibnamefont {Hatsuda}},\ }\href
  {\doibase 10.1103/PhysRevLett.99.022001} {\bibfield  {journal} {\bibinfo
  {journal} {Phys. Rev. Lett.}\ }\textbf {\bibinfo {volume} {99}},\ \bibinfo
  {pages} {022001} (\bibinfo {year} {2007})}\BibitemShut {NoStop}%
\bibitem [{\citenamefont {Aoki}\ \emph {et~al.}(2010)\citenamefont {Aoki},
  \citenamefont {Hatsuda},\ and\ \citenamefont {Ishii}}]{Aoki:2009ji}%
  \BibitemOpen
  \bibfield  {author} {\bibinfo {author} {\bibfnamefont {S.}~\bibnamefont
  {Aoki}}, \bibinfo {author} {\bibfnamefont {T.}~\bibnamefont {Hatsuda}}, \
  and\ \bibinfo {author} {\bibfnamefont {N.}~\bibnamefont {Ishii}},\ }\href
  {\doibase 10.1143/PTP.123.89} {\bibfield  {journal} {\bibinfo  {journal}
  {Progress of Theoretical Physics}\ }\textbf {\bibinfo {volume} {123}},\
  \bibinfo {pages} {89} (\bibinfo {year} {2010})}\BibitemShut {NoStop}%
\bibitem [{\citenamefont {Aoki}\ \emph {et~al.}(2012)\citenamefont {Aoki},
  \citenamefont {Doi}, \citenamefont {Hatsuda}, \citenamefont {Ikeda},
  \citenamefont {Inoue}, \citenamefont {Ishii}, \citenamefont {Murano},
  \citenamefont {Nemura},\ and\ \citenamefont {Sasaki}}]{Aoki:2012tk}%
  \BibitemOpen
  \bibfield  {author} {\bibinfo {author} {\bibfnamefont {S.}~\bibnamefont
  {Aoki}}, \bibinfo {author} {\bibfnamefont {T.}~\bibnamefont {Doi}}, \bibinfo
  {author} {\bibfnamefont {T.}~\bibnamefont {Hatsuda}}, \bibinfo {author}
  {\bibfnamefont {Y.}~\bibnamefont {Ikeda}}, \bibinfo {author} {\bibfnamefont
  {T.}~\bibnamefont {Inoue}}, \bibinfo {author} {\bibfnamefont
  {N.}~\bibnamefont {Ishii}}, \bibinfo {author} {\bibfnamefont
  {K.}~\bibnamefont {Murano}}, \bibinfo {author} {\bibfnamefont
  {H.}~\bibnamefont {Nemura}}, \ and\ \bibinfo {author} {\bibfnamefont
  {K.}~\bibnamefont {Sasaki}} (\bibinfo {collaboration} {HAL QCD}),\ }\href
  {\doibase 10.1093/ptep/pts010} {\bibfield  {journal} {\bibinfo  {journal}
  {PTEP}\ }\textbf {\bibinfo {volume} {2012}},\ \bibinfo {pages} {01A105}
  (\bibinfo {year} {2012})}\BibitemShut {NoStop}%
\bibitem [{\citenamefont {Aoki}\ \emph {et~al.}(2013)\citenamefont {Aoki},
  \citenamefont {Ishii}, \citenamefont {Doi}, \citenamefont {Ikeda},\ and\
  \citenamefont {Inoue}}]{Aoki2013PRD}%
  \BibitemOpen
  \bibfield  {author} {\bibinfo {author} {\bibfnamefont {S.}~\bibnamefont
  {Aoki}}, \bibinfo {author} {\bibfnamefont {N.}~\bibnamefont {Ishii}},
  \bibinfo {author} {\bibfnamefont {T.}~\bibnamefont {Doi}}, \bibinfo {author}
  {\bibfnamefont {Y.}~\bibnamefont {Ikeda}}, \ and\ \bibinfo {author}
  {\bibfnamefont {T.}~\bibnamefont {Inoue}},\ }\href {\doibase
  10.1103/PhysRevD.88.014036} {\bibfield  {journal} {\bibinfo  {journal} {Phys.
  Rev. D}\ }\textbf {\bibinfo {volume} {88}},\ \bibinfo {pages} {014036}
  (\bibinfo {year} {2013})}\BibitemShut {NoStop}%
\bibitem [{\citenamefont {Doi}\ \emph {et~al.}(2012)\citenamefont {Doi},
  \citenamefont {Aoki}, \citenamefont {Hatsuda}, \citenamefont {Ikeda},
  \citenamefont {Inoue}, \citenamefont {Ishii}, \citenamefont {Murano},
  \citenamefont {Nemura}, \citenamefont {Sasaki},\ and\ \citenamefont {{HAL QCD
  Collaboration}}}]{Doi:2011gq}%
  \BibitemOpen
  \bibfield  {author} {\bibinfo {author} {\bibfnamefont {T.}~\bibnamefont
  {Doi}}, \bibinfo {author} {\bibfnamefont {S.}~\bibnamefont {Aoki}}, \bibinfo
  {author} {\bibfnamefont {T.}~\bibnamefont {Hatsuda}}, \bibinfo {author}
  {\bibfnamefont {Y.}~\bibnamefont {Ikeda}}, \bibinfo {author} {\bibfnamefont
  {T.}~\bibnamefont {Inoue}}, \bibinfo {author} {\bibfnamefont
  {N.}~\bibnamefont {Ishii}}, \bibinfo {author} {\bibfnamefont
  {K.}~\bibnamefont {Murano}}, \bibinfo {author} {\bibfnamefont
  {H.}~\bibnamefont {Nemura}}, \bibinfo {author} {\bibfnamefont
  {K.}~\bibnamefont {Sasaki}}, \ and\ \bibinfo {author} {\bibnamefont {{HAL QCD
  Collaboration}}},\ }\href {\doibase 10.1143/PTP.127.723} {\bibfield
  {journal} {\bibinfo  {journal} {Progress of Theoretical Physics}\ }\textbf
  {\bibinfo {volume} {127}},\ \bibinfo {pages} {723} (\bibinfo {year}
  {2012})}\BibitemShut {NoStop}%
\bibitem [{\citenamefont {Ishii}\ \emph {et~al.}(2012)\citenamefont {Ishii},
  \citenamefont {Aoki}, \citenamefont {Doi}, \citenamefont {Hatsuda},
  \citenamefont {Ikeda}, \citenamefont {Inoue}, \citenamefont {Murano},
  \citenamefont {Nemura},\ and\ \citenamefont {Sasaki}}]{HALQCD:2012aa}%
  \BibitemOpen
  \bibfield  {author} {\bibinfo {author} {\bibfnamefont {N.}~\bibnamefont
  {Ishii}}, \bibinfo {author} {\bibfnamefont {S.}~\bibnamefont {Aoki}},
  \bibinfo {author} {\bibfnamefont {T.}~\bibnamefont {Doi}}, \bibinfo {author}
  {\bibfnamefont {T.}~\bibnamefont {Hatsuda}}, \bibinfo {author} {\bibfnamefont
  {Y.}~\bibnamefont {Ikeda}}, \bibinfo {author} {\bibfnamefont
  {T.}~\bibnamefont {Inoue}}, \bibinfo {author} {\bibfnamefont
  {K.}~\bibnamefont {Murano}}, \bibinfo {author} {\bibfnamefont
  {H.}~\bibnamefont {Nemura}}, \ and\ \bibinfo {author} {\bibfnamefont
  {K.}~\bibnamefont {Sasaki}},\ }\href {\doibase
  http://dx.doi.org/10.1016/j.physletb.2012.04.076} {\bibfield  {journal}
  {\bibinfo  {journal} {Physics Letters B}\ }\textbf {\bibinfo {volume}
  {712}},\ \bibinfo {pages} {437 } (\bibinfo {year} {2012})}\BibitemShut
  {NoStop}%
\bibitem [{\citenamefont {Doi}\ and\ \citenamefont
  {Endres}(2013)}]{Doi:2012xd}%
  \BibitemOpen
  \bibfield  {author} {\bibinfo {author} {\bibfnamefont {T.}~\bibnamefont
  {Doi}}\ and\ \bibinfo {author} {\bibfnamefont {M.~G.}\ \bibnamefont
  {Endres}},\ }\href {\doibase http://dx.doi.org/10.1016/j.cpc.2012.09.004}
  {\bibfield  {journal} {\bibinfo  {journal} {Computer Physics Communications}\
  }\textbf {\bibinfo {volume} {184}},\ \bibinfo {pages} {117 } (\bibinfo {year}
  {2013})}\BibitemShut {NoStop}%
\bibitem [{\citenamefont {Detmold}\ and\ \citenamefont
  {Orginos}(2013)}]{Detmold:2012eu}%
  \BibitemOpen
  \bibfield  {author} {\bibinfo {author} {\bibfnamefont {W.}~\bibnamefont
  {Detmold}}\ and\ \bibinfo {author} {\bibfnamefont {K.}~\bibnamefont
  {Orginos}},\ }\href {\doibase 10.1103/PhysRevD.87.114512} {\bibfield
  {journal} {\bibinfo  {journal} {Phys. Rev. D}\ }\textbf {\bibinfo {volume}
  {87}},\ \bibinfo {pages} {114512} (\bibinfo {year} {2013})}\BibitemShut
  {NoStop}%
\bibitem [{\citenamefont {G\"unther}\ \emph {et~al.}(2013)\citenamefont
  {G\"unther}, \citenamefont {T\'oth},\ and\ \citenamefont
  {Varnhorst}}]{Gunther:2013xj}%
  \BibitemOpen
  \bibfield  {author} {\bibinfo {author} {\bibfnamefont {J.}~\bibnamefont
  {G\"unther}}, \bibinfo {author} {\bibfnamefont {B.~C.}\ \bibnamefont
  {T\'oth}}, \ and\ \bibinfo {author} {\bibfnamefont {L.}~\bibnamefont
  {Varnhorst}},\ }\href {\doibase 10.1103/PhysRevD.87.094513} {\bibfield
  {journal} {\bibinfo  {journal} {Phys. Rev. D}\ }\textbf {\bibinfo {volume}
  {87}},\ \bibinfo {pages} {094513} (\bibinfo {year} {2013})}\BibitemShut
  {NoStop}%
\bibitem [{\citenamefont {Nemura}(2016)}]{Nemura2016CPC}%
  \BibitemOpen
  \bibfield  {author} {\bibinfo {author} {\bibfnamefont {H.}~\bibnamefont
  {Nemura}},\ }\href {\doibase http://dx.doi.org/10.1016/j.cpc.2016.05.014}
  {\bibfield  {journal} {\bibinfo  {journal} {Computer Physics Communications}\
  }\textbf {\bibinfo {volume} {207}},\ \bibinfo {pages} {91 } (\bibinfo {year}
  {2016})}\BibitemShut {NoStop}%
\bibitem [{\citenamefont {Barnea}\ \emph {et~al.}(2015)\citenamefont {Barnea},
  \citenamefont {Contessi}, \citenamefont {Gazit}, \citenamefont {Pederiva},\
  and\ \citenamefont {van Kolck}}]{Barnea2005lqcd}%
  \BibitemOpen
  \bibfield  {author} {\bibinfo {author} {\bibfnamefont {N.}~\bibnamefont
  {Barnea}}, \bibinfo {author} {\bibfnamefont {L.}~\bibnamefont {Contessi}},
  \bibinfo {author} {\bibfnamefont {D.}~\bibnamefont {Gazit}}, \bibinfo
  {author} {\bibfnamefont {F.}~\bibnamefont {Pederiva}}, \ and\ \bibinfo
  {author} {\bibfnamefont {U.}~\bibnamefont {van Kolck}},\ }\href {\doibase
  10.1103/PhysRevLett.114.052501} {\bibfield  {journal} {\bibinfo  {journal}
  {Phys. Rev. Lett.}\ }\textbf {\bibinfo {volume} {114}},\ \bibinfo {pages}
  {052501} (\bibinfo {year} {2015})}\BibitemShut {NoStop}%
\bibitem [{\citenamefont {Contessi}\ \emph {et~al.}(2017)\citenamefont
  {Contessi}, \citenamefont {Lovato}, \citenamefont {Pederiva}, \citenamefont
  {Roggero}, \citenamefont {Kirscher},\ and\ \citenamefont {van
  Kolck}}]{Contessi2017pilessEFT}%
  \BibitemOpen
  \bibfield  {author} {\bibinfo {author} {\bibfnamefont {L.}~\bibnamefont
  {Contessi}}, \bibinfo {author} {\bibfnamefont {A.}~\bibnamefont {Lovato}},
  \bibinfo {author} {\bibfnamefont {F.}~\bibnamefont {Pederiva}}, \bibinfo
  {author} {\bibfnamefont {A.}~\bibnamefont {Roggero}}, \bibinfo {author}
  {\bibfnamefont {J.}~\bibnamefont {Kirscher}}, \ and\ \bibinfo {author}
  {\bibfnamefont {U.}~\bibnamefont {van Kolck}},\ }\href {\doibase
  https://doi.org/10.1016/j.physletb.2017.07.048} {\bibfield  {journal}
  {\bibinfo  {journal} {Physics Letters B}\ }\textbf {\bibinfo {volume}
  {772}},\ \bibinfo {pages} {839 } (\bibinfo {year} {2017})}\BibitemShut
  {NoStop}%
\bibitem [{\citenamefont {Inoue}\ \emph {et~al.}(2011)\citenamefont {Inoue},
  \citenamefont {Ishii}, \citenamefont {Aoki}, \citenamefont {Doi},
  \citenamefont {Hatsuda}, \citenamefont {Ikeda}, \citenamefont {Murano},
  \citenamefont {Nemura},\ and\ \citenamefont {Sasaki}}]{Inoue:2010es}%
  \BibitemOpen
  \bibfield  {author} {\bibinfo {author} {\bibfnamefont {T.}~\bibnamefont
  {Inoue}}, \bibinfo {author} {\bibfnamefont {N.}~\bibnamefont {Ishii}},
  \bibinfo {author} {\bibfnamefont {S.}~\bibnamefont {Aoki}}, \bibinfo {author}
  {\bibfnamefont {T.}~\bibnamefont {Doi}}, \bibinfo {author} {\bibfnamefont
  {T.}~\bibnamefont {Hatsuda}}, \bibinfo {author} {\bibfnamefont
  {Y.}~\bibnamefont {Ikeda}}, \bibinfo {author} {\bibfnamefont
  {K.}~\bibnamefont {Murano}}, \bibinfo {author} {\bibfnamefont
  {H.}~\bibnamefont {Nemura}}, \ and\ \bibinfo {author} {\bibfnamefont
  {K.}~\bibnamefont {Sasaki}} (\bibinfo {collaboration} {HAL QCD
  Collaboration}),\ }\href {\doibase 10.1103/PhysRevLett.106.162002} {\bibfield
   {journal} {\bibinfo  {journal} {Phys. Rev. Lett.}\ }\textbf {\bibinfo
  {volume} {106}},\ \bibinfo {pages} {162002} (\bibinfo {year}
  {2011})}\BibitemShut {NoStop}%
\bibitem [{\citenamefont {Inoue}\ \emph {et~al.}(2012)\citenamefont {Inoue},
  \citenamefont {Aoki}, \citenamefont {Doi}, \citenamefont {Hatsuda},
  \citenamefont {Ikeda}, \citenamefont {Ishii}, \citenamefont {Murano},
  \citenamefont {Nemura},\ and\ \citenamefont {Sasaki}}]{Inoue:2011ai}%
  \BibitemOpen
  \bibfield  {author} {\bibinfo {author} {\bibfnamefont {T.}~\bibnamefont
  {Inoue}}, \bibinfo {author} {\bibfnamefont {S.}~\bibnamefont {Aoki}},
  \bibinfo {author} {\bibfnamefont {T.}~\bibnamefont {Doi}}, \bibinfo {author}
  {\bibfnamefont {T.}~\bibnamefont {Hatsuda}}, \bibinfo {author} {\bibfnamefont
  {Y.}~\bibnamefont {Ikeda}}, \bibinfo {author} {\bibfnamefont
  {N.}~\bibnamefont {Ishii}}, \bibinfo {author} {\bibfnamefont
  {K.}~\bibnamefont {Murano}}, \bibinfo {author} {\bibfnamefont
  {H.}~\bibnamefont {Nemura}}, \ and\ \bibinfo {author} {\bibfnamefont
  {K.}~\bibnamefont {Sasaki}},\ }\href {\doibase
  http://dx.doi.org/10.1016/j.nuclphysa.2012.02.008} {\bibfield  {journal}
  {\bibinfo  {journal} {Nuclear Physics A}\ }\textbf {\bibinfo {volume}
  {881}},\ \bibinfo {pages} {28 } (\bibinfo {year} {2012})}\BibitemShut
  {NoStop}%
\bibitem [{\citenamefont {Inoue}\ \emph {et~al.}(2013)\citenamefont {Inoue},
  \citenamefont {Aoki}, \citenamefont {Doi}, \citenamefont {Hatsuda},
  \citenamefont {Ikeda}, \citenamefont {Ishii}, \citenamefont {Murano},
  \citenamefont {Nemura},\ and\ \citenamefont {Sasaki}}]{Inoue:2013nfe}%
  \BibitemOpen
  \bibfield  {author} {\bibinfo {author} {\bibfnamefont {T.}~\bibnamefont
  {Inoue}}, \bibinfo {author} {\bibfnamefont {S.}~\bibnamefont {Aoki}},
  \bibinfo {author} {\bibfnamefont {T.}~\bibnamefont {Doi}}, \bibinfo {author}
  {\bibfnamefont {T.}~\bibnamefont {Hatsuda}}, \bibinfo {author} {\bibfnamefont
  {Y.}~\bibnamefont {Ikeda}}, \bibinfo {author} {\bibfnamefont
  {N.}~\bibnamefont {Ishii}}, \bibinfo {author} {\bibfnamefont
  {K.}~\bibnamefont {Murano}}, \bibinfo {author} {\bibfnamefont
  {H.}~\bibnamefont {Nemura}}, \ and\ \bibinfo {author} {\bibfnamefont
  {K.}~\bibnamefont {Sasaki}} (\bibinfo {collaboration} {HAL QCD
  Collaboration}),\ }\href {\doibase 10.1103/PhysRevLett.111.112503} {\bibfield
   {journal} {\bibinfo  {journal} {Phys. Rev. Lett.}\ }\textbf {\bibinfo
  {volume} {111}},\ \bibinfo {pages} {112503} (\bibinfo {year}
  {2013})}\BibitemShut {NoStop}%
\bibitem [{\citenamefont {Inoue}\ \emph {et~al.}(2015)\citenamefont {Inoue},
  \citenamefont {Aoki}, \citenamefont {Charron}, \citenamefont {Doi},
  \citenamefont {Hatsuda}, \citenamefont {Ikeda}, \citenamefont {Ishii},
  \citenamefont {Murano}, \citenamefont {Nemura},\ and\ \citenamefont
  {Sasaki}}]{Inoue:2014ipa}%
  \BibitemOpen
  \bibfield  {author} {\bibinfo {author} {\bibfnamefont {T.}~\bibnamefont
  {Inoue}}, \bibinfo {author} {\bibfnamefont {S.}~\bibnamefont {Aoki}},
  \bibinfo {author} {\bibfnamefont {B.}~\bibnamefont {Charron}}, \bibinfo
  {author} {\bibfnamefont {T.}~\bibnamefont {Doi}}, \bibinfo {author}
  {\bibfnamefont {T.}~\bibnamefont {Hatsuda}}, \bibinfo {author} {\bibfnamefont
  {Y.}~\bibnamefont {Ikeda}}, \bibinfo {author} {\bibfnamefont
  {N.}~\bibnamefont {Ishii}}, \bibinfo {author} {\bibfnamefont
  {K.}~\bibnamefont {Murano}}, \bibinfo {author} {\bibfnamefont
  {H.}~\bibnamefont {Nemura}}, \ and\ \bibinfo {author} {\bibfnamefont
  {K.}~\bibnamefont {Sasaki}} (\bibinfo {collaboration} {HAL QCD
  Collaboration}),\ }\href {\doibase 10.1103/PhysRevC.91.011001} {\bibfield
  {journal} {\bibinfo  {journal} {Phys. Rev. C}\ }\textbf {\bibinfo {volume}
  {91}},\ \bibinfo {pages} {011001} (\bibinfo {year} {2015})}\BibitemShut
  {NoStop}%
\bibitem [{\citenamefont {Som\`a}\ \emph
  {et~al.}(2014{\natexlab{a}})\citenamefont {Som\`a}, \citenamefont
  {Cipollone}, \citenamefont {Barbieri}, \citenamefont {Navr\'atil},\ and\
  \citenamefont {Duguet}}]{Soma2014s2n}%
  \BibitemOpen
  \bibfield  {author} {\bibinfo {author} {\bibfnamefont {V.}~\bibnamefont
  {Som\`a}}, \bibinfo {author} {\bibfnamefont {A.}~\bibnamefont {Cipollone}},
  \bibinfo {author} {\bibfnamefont {C.}~\bibnamefont {Barbieri}}, \bibinfo
  {author} {\bibfnamefont {P.}~\bibnamefont {Navr\'atil}}, \ and\ \bibinfo
  {author} {\bibfnamefont {T.}~\bibnamefont {Duguet}},\ }\href {\doibase
  10.1103/PhysRevC.89.061301} {\bibfield  {journal} {\bibinfo  {journal} {Phys.
  Rev. C}\ }\textbf {\bibinfo {volume} {89}},\ \bibinfo {pages} {061301}
  (\bibinfo {year} {2014}{\natexlab{a}})}\BibitemShut {NoStop}%
\bibitem [{\citenamefont {Hergert}\ \emph {et~al.}(2014)\citenamefont
  {Hergert}, \citenamefont {Bogner}, \citenamefont {Morris}, \citenamefont
  {Binder}, \citenamefont {Calci}, \citenamefont {Langhammer},\ and\
  \citenamefont {Roth}}]{Hergert2014Ni}%
  \BibitemOpen
  \bibfield  {author} {\bibinfo {author} {\bibfnamefont {H.}~\bibnamefont
  {Hergert}}, \bibinfo {author} {\bibfnamefont {S.~K.}\ \bibnamefont {Bogner}},
  \bibinfo {author} {\bibfnamefont {T.~D.}\ \bibnamefont {Morris}}, \bibinfo
  {author} {\bibfnamefont {S.}~\bibnamefont {Binder}}, \bibinfo {author}
  {\bibfnamefont {A.}~\bibnamefont {Calci}}, \bibinfo {author} {\bibfnamefont
  {J.}~\bibnamefont {Langhammer}}, \ and\ \bibinfo {author} {\bibfnamefont
  {R.}~\bibnamefont {Roth}},\ }\href {\doibase 10.1103/PhysRevC.90.041302}
  {\bibfield  {journal} {\bibinfo  {journal} {Phys. Rev. C}\ }\textbf {\bibinfo
  {volume} {90}},\ \bibinfo {pages} {041302} (\bibinfo {year}
  {2014})}\BibitemShut {NoStop}%
\bibitem [{\citenamefont {Hagen}\ \emph {et~al.}(2014)\citenamefont {Hagen},
  \citenamefont {Papenbrock}, \citenamefont {Hjorth-Jensen},\ and\
  \citenamefont {Dean}}]{Hagen2014CCMrev}%
  \BibitemOpen
  \bibfield  {author} {\bibinfo {author} {\bibfnamefont {G.}~\bibnamefont
  {Hagen}}, \bibinfo {author} {\bibfnamefont {T.}~\bibnamefont {Papenbrock}},
  \bibinfo {author} {\bibfnamefont {M.}~\bibnamefont {Hjorth-Jensen}}, \ and\
  \bibinfo {author} {\bibfnamefont {D.~J.}\ \bibnamefont {Dean}},\ }\href
  {http://stacks.iop.org/0034-4885/77/i=9/a=096302} {\bibfield  {journal}
  {\bibinfo  {journal} {Rep. Prog. Phys.}\ }\textbf {\bibinfo {volume} {77}},\
  \bibinfo {pages} {096302} (\bibinfo {year} {2014})}\BibitemShut {NoStop}%
\bibitem [{\citenamefont {Dickhoff}\ and\ \citenamefont
  {Barbieri}(2004)}]{Dickhoff2004ppnp}%
  \BibitemOpen
  \bibfield  {author} {\bibinfo {author} {\bibfnamefont {W.}~\bibnamefont
  {Dickhoff}}\ and\ \bibinfo {author} {\bibfnamefont {C.}~\bibnamefont
  {Barbieri}},\ }\href {\doibase http://dx.doi.org/10.1016/j.ppnp.2004.02.038}
  {\bibfield  {journal} {\bibinfo  {journal} {Progress in Particle and Nuclear
  Physics}\ }\textbf {\bibinfo {volume} {52}},\ \bibinfo {pages} {377 }
  (\bibinfo {year} {2004})}\BibitemShut {NoStop}%
\bibitem [{\citenamefont {Carbone}\ \emph {et~al.}(2013)\citenamefont
  {Carbone}, \citenamefont {Cipollone}, \citenamefont {Barbieri}, \citenamefont
  {Rios},\ and\ \citenamefont {Polls}}]{Carbone2013tnf}%
  \BibitemOpen
  \bibfield  {author} {\bibinfo {author} {\bibfnamefont {A.}~\bibnamefont
  {Carbone}}, \bibinfo {author} {\bibfnamefont {A.}~\bibnamefont {Cipollone}},
  \bibinfo {author} {\bibfnamefont {C.}~\bibnamefont {Barbieri}}, \bibinfo
  {author} {\bibfnamefont {A.}~\bibnamefont {Rios}}, \ and\ \bibinfo {author}
  {\bibfnamefont {A.}~\bibnamefont {Polls}},\ }\href {\doibase
  10.1103/PhysRevC.88.054326} {\bibfield  {journal} {\bibinfo  {journal} {Phys.
  Rev. C}\ }\textbf {\bibinfo {volume} {88}},\ \bibinfo {pages} {054326}
  (\bibinfo {year} {2013})}\BibitemShut {NoStop}%
\bibitem [{\citenamefont {Hebeler}\ \emph {et~al.}(2015)\citenamefont
  {Hebeler}, \citenamefont {Holt}, \citenamefont {Men\'endez},\ and\
  \citenamefont {Schwenk}}]{Hebeler2015ChiralRev}%
  \BibitemOpen
  \bibfield  {author} {\bibinfo {author} {\bibfnamefont {K.}~\bibnamefont
  {Hebeler}}, \bibinfo {author} {\bibfnamefont {J.}~\bibnamefont {Holt}},
  \bibinfo {author} {\bibfnamefont {J.}~\bibnamefont {Men\'endez}}, \ and\
  \bibinfo {author} {\bibfnamefont {A.}~\bibnamefont {Schwenk}},\ }\href
  {\doibase 10.1146/annurev-nucl-102313-025446} {\bibfield  {journal} {\bibinfo
   {journal} {Annual Review of Nuclear and Particle Science}\ }\textbf
  {\bibinfo {volume} {65}},\ \bibinfo {pages} {457} (\bibinfo {year}
  {2015})}\BibitemShut {NoStop}%
\bibitem [{\citenamefont {Cipollone}\ \emph {et~al.}(2013)\citenamefont
  {Cipollone}, \citenamefont {Barbieri},\ and\ \citenamefont
  {Navr\'atil}}]{Cipollone2013prl}%
  \BibitemOpen
  \bibfield  {author} {\bibinfo {author} {\bibfnamefont {A.}~\bibnamefont
  {Cipollone}}, \bibinfo {author} {\bibfnamefont {C.}~\bibnamefont {Barbieri}},
  \ and\ \bibinfo {author} {\bibfnamefont {P.}~\bibnamefont {Navr\'atil}},\
  }\href {\doibase 10.1103/PhysRevLett.111.062501} {\bibfield  {journal}
  {\bibinfo  {journal} {Phys. Rev. Lett.}\ }\textbf {\bibinfo {volume} {111}},\
  \bibinfo {pages} {062501} (\bibinfo {year} {2013})}\BibitemShut {NoStop}%
\bibitem [{\citenamefont {Ekstr\"om}\ \emph {et~al.}(2015)\citenamefont
  {Ekstr\"om}, \citenamefont {Jansen}, \citenamefont {Wendt}, \citenamefont
  {Hagen}, \citenamefont {Papenbrock}, \citenamefont {Carlsson}, \citenamefont
  {Forss\'en}, \citenamefont {Hjorth-Jensen}, \citenamefont {Navr\'atil},\ and\
  \citenamefont {Nazarewicz}}]{Ekstrom2015NNLOsat}%
  \BibitemOpen
  \bibfield  {author} {\bibinfo {author} {\bibfnamefont {A.}~\bibnamefont
  {Ekstr\"om}}, \bibinfo {author} {\bibfnamefont {G.~R.}\ \bibnamefont
  {Jansen}}, \bibinfo {author} {\bibfnamefont {K.~A.}\ \bibnamefont {Wendt}},
  \bibinfo {author} {\bibfnamefont {G.}~\bibnamefont {Hagen}}, \bibinfo
  {author} {\bibfnamefont {T.}~\bibnamefont {Papenbrock}}, \bibinfo {author}
  {\bibfnamefont {B.~D.}\ \bibnamefont {Carlsson}}, \bibinfo {author}
  {\bibfnamefont {C.}~\bibnamefont {Forss\'en}}, \bibinfo {author}
  {\bibfnamefont {M.}~\bibnamefont {Hjorth-Jensen}}, \bibinfo {author}
  {\bibfnamefont {P.}~\bibnamefont {Navr\'atil}}, \ and\ \bibinfo {author}
  {\bibfnamefont {W.}~\bibnamefont {Nazarewicz}},\ }\href {\doibase
  10.1103/PhysRevC.91.051301} {\bibfield  {journal} {\bibinfo  {journal} {Phys.
  Rev. C}\ }\textbf {\bibinfo {volume} {91}},\ \bibinfo {pages} {051301}
  (\bibinfo {year} {2015})}\BibitemShut {NoStop}%
\bibitem [{\citenamefont {Lapoux}\ \emph {et~al.}(2016)\citenamefont {Lapoux},
  \citenamefont {Som\`a}, \citenamefont {Barbieri}, \citenamefont {Hergert},
  \citenamefont {Holt},\ and\ \citenamefont {Stroberg}}]{Lapoux2016prlOx}%
  \BibitemOpen
  \bibfield  {author} {\bibinfo {author} {\bibfnamefont {V.}~\bibnamefont
  {Lapoux}}, \bibinfo {author} {\bibfnamefont {V.}~\bibnamefont {Som\`a}},
  \bibinfo {author} {\bibfnamefont {C.}~\bibnamefont {Barbieri}}, \bibinfo
  {author} {\bibfnamefont {H.}~\bibnamefont {Hergert}}, \bibinfo {author}
  {\bibfnamefont {J.~D.}\ \bibnamefont {Holt}}, \ and\ \bibinfo {author}
  {\bibfnamefont {S.~R.}\ \bibnamefont {Stroberg}},\ }\href {\doibase
  10.1103/PhysRevLett.117.052501} {\bibfield  {journal} {\bibinfo  {journal}
  {Phys. Rev. Lett.}\ }\textbf {\bibinfo {volume} {117}},\ \bibinfo {pages}
  {052501} (\bibinfo {year} {2016})}\BibitemShut {NoStop}%
\bibitem [{\citenamefont {Garcia~Ruiz}\ \emph {et~al.}(2016)\citenamefont
  {Garcia~Ruiz}, \citenamefont {Bissell}, \citenamefont {Blaum}, \citenamefont
  {Ekstrom}, \citenamefont {Frommgen}, \citenamefont {Hagen}, \citenamefont
  {Hammen}, \citenamefont {Hebeler}, \citenamefont {Holt}, \citenamefont
  {Jansen}, \citenamefont {Kowalska}, \citenamefont {Kreim}, \citenamefont
  {Nazarewicz}, \citenamefont {Neugart}, \citenamefont {Neyens}, \citenamefont
  {Nortershauser}, \citenamefont {Papenbrock}, \citenamefont {Papuga},
  \citenamefont {Schwenk}, \citenamefont {Simonis}, \citenamefont {Wendt},\
  and\ \citenamefont {Yordanov}}]{Ruiz2016NatCa48}%
  \BibitemOpen
  \bibfield  {author} {\bibinfo {author} {\bibfnamefont {R.~F.}\ \bibnamefont
  {Garcia~Ruiz}}, \bibinfo {author} {\bibfnamefont {M.~L.}\ \bibnamefont
  {Bissell}}, \bibinfo {author} {\bibfnamefont {K.}~\bibnamefont {Blaum}},
  \bibinfo {author} {\bibfnamefont {A.}~\bibnamefont {Ekstrom}}, \bibinfo
  {author} {\bibfnamefont {N.}~\bibnamefont {Frommgen}}, \bibinfo {author}
  {\bibfnamefont {G.}~\bibnamefont {Hagen}}, \bibinfo {author} {\bibfnamefont
  {M.}~\bibnamefont {Hammen}}, \bibinfo {author} {\bibfnamefont
  {K.}~\bibnamefont {Hebeler}}, \bibinfo {author} {\bibfnamefont {J.~D.}\
  \bibnamefont {Holt}}, \bibinfo {author} {\bibfnamefont {G.~R.}\ \bibnamefont
  {Jansen}}, \bibinfo {author} {\bibfnamefont {M.}~\bibnamefont {Kowalska}},
  \bibinfo {author} {\bibfnamefont {K.}~\bibnamefont {Kreim}}, \bibinfo
  {author} {\bibfnamefont {W.}~\bibnamefont {Nazarewicz}}, \bibinfo {author}
  {\bibfnamefont {R.}~\bibnamefont {Neugart}}, \bibinfo {author} {\bibfnamefont
  {G.}~\bibnamefont {Neyens}}, \bibinfo {author} {\bibfnamefont
  {W.}~\bibnamefont {Nortershauser}}, \bibinfo {author} {\bibfnamefont
  {T.}~\bibnamefont {Papenbrock}}, \bibinfo {author} {\bibfnamefont
  {J.}~\bibnamefont {Papuga}}, \bibinfo {author} {\bibfnamefont
  {A.}~\bibnamefont {Schwenk}}, \bibinfo {author} {\bibfnamefont
  {J.}~\bibnamefont {Simonis}}, \bibinfo {author} {\bibfnamefont {K.~A.}\
  \bibnamefont {Wendt}}, \ and\ \bibinfo {author} {\bibfnamefont {D.~T.}\
  \bibnamefont {Yordanov}},\ }\href {http://dx.doi.org/10.1038/nphys3645}
  {\bibfield  {journal} {\bibinfo  {journal} {Nat Phys}\ }\textbf {\bibinfo
  {volume} {12}},\ \bibinfo {pages} {594} (\bibinfo {year} {2016})}\BibitemShut
  {NoStop}%
\bibitem [{\citenamefont {Barbieri}\ and\ \citenamefont
  {Dickhoff}(2002)}]{Barbieri2002o16}%
  \BibitemOpen
  \bibfield  {author} {\bibinfo {author} {\bibfnamefont {C.}~\bibnamefont
  {Barbieri}}\ and\ \bibinfo {author} {\bibfnamefont {W.~H.}\ \bibnamefont
  {Dickhoff}},\ }\href {\doibase 10.1103/PhysRevC.65.064313} {\bibfield
  {journal} {\bibinfo  {journal} {Phys. Rev. C}\ }\textbf {\bibinfo {volume}
  {65}},\ \bibinfo {pages} {064313} (\bibinfo {year} {2002})}\BibitemShut
  {NoStop}%
\bibitem [{\citenamefont {Barbieri}(2006)}]{Barbieri2006plbO16}%
  \BibitemOpen
  \bibfield  {author} {\bibinfo {author} {\bibfnamefont {C.}~\bibnamefont
  {Barbieri}},\ }\href@noop {} {\bibfield  {journal} {\bibinfo  {journal}
  {Phys. Lett. B}\ }\textbf {\bibinfo {volume} {643}},\ \bibinfo {pages} {268}
  (\bibinfo {year} {2006})}\BibitemShut {NoStop}%
\bibitem [{\citenamefont {Barbieri}\ and\ \citenamefont
  {Hjorth-Jensen}(2009)}]{Barbieri2009Ni56prc}%
  \BibitemOpen
  \bibfield  {author} {\bibinfo {author} {\bibfnamefont {C.}~\bibnamefont
  {Barbieri}}\ and\ \bibinfo {author} {\bibfnamefont {M.}~\bibnamefont
  {Hjorth-Jensen}},\ }\href {\doibase 10.1103/PhysRevC.79.064313} {\bibfield
  {journal} {\bibinfo  {journal} {Phys. Rev. C}\ }\textbf {\bibinfo {volume}
  {79}},\ \bibinfo {pages} {064313} (\bibinfo {year} {2009})}\BibitemShut
  {NoStop}%
\bibitem [{\citenamefont {Doi}\ \emph {et~al.}(2016)\citenamefont {Doi} \emph
  {et~al.}}]{Doi:2015oha}%
  \BibitemOpen
  \bibfield  {author} {\bibinfo {author} {\bibfnamefont {T.}~\bibnamefont
  {Doi}} \emph {et~al.},\ }\bibfield  {booktitle} {\emph {\bibinfo {booktitle}
  {{Proceedings, 33rd International Symposium on Lattice Field Theory (Lattice
  2015): Kobe, Japan, July 14-18, 2015}}},\ }\href@noop {} {\bibfield
  {journal} {\bibinfo  {journal} {PoS LATTICE}\ }\textbf {\bibinfo {volume}
  {2015}},\ \bibinfo {pages} {086} (\bibinfo {year} {2016})},\ \Eprint
  {http://arxiv.org/abs/1512.01610} {arXiv:1512.01610 [hep-lat]} \BibitemShut
  {NoStop}%
\bibitem [{\citenamefont {Ishii}\ \emph {et~al.}(2016)\citenamefont {Ishii}
  \emph {et~al.}}]{Ishii:2016zsf}%
  \BibitemOpen
  \bibfield  {author} {\bibinfo {author} {\bibfnamefont {N.}~\bibnamefont
  {Ishii}} \emph {et~al.},\ }\bibfield  {booktitle} {\emph {\bibinfo
  {booktitle} {{Proceedings, 33rd International Symposium on Lattice Field
  Theory (Lattice 2015): Kobe, Japan, July 14-18, 2015}}},\ }\href@noop {}
  {\bibfield  {journal} {\bibinfo  {journal} {PoS LATTICE}\ }\textbf {\bibinfo
  {volume} {2015}},\ \bibinfo {pages} {087} (\bibinfo {year}
  {2016})}\BibitemShut {NoStop}%
\bibitem [{\citenamefont {Sasaki}\ \emph {et~al.}(2016)\citenamefont {Sasaki}
  \emph {et~al.}}]{Sasaki:2016gpc}%
  \BibitemOpen
  \bibfield  {author} {\bibinfo {author} {\bibfnamefont {K.}~\bibnamefont
  {Sasaki}} \emph {et~al.},\ }\bibfield  {booktitle} {\emph {\bibinfo
  {booktitle} {{Proceedings, 33rd International Symposium on Lattice Field
  Theory (Lattice 2015): Kobe, Japan, July 14-18, 2015}}},\ }\href@noop {}
  {\bibfield  {journal} {\bibinfo  {journal} {PoS LATTICE}\ }\textbf {\bibinfo
  {volume} {2015}},\ \bibinfo {pages} {088} (\bibinfo {year}
  {2016})}\BibitemShut {NoStop}%
\bibitem [{\citenamefont {Nemura}\ \emph {et~al.}(2016)\citenamefont {Nemura}
  \emph {et~al.}}]{Nemura:2016sty}%
  \BibitemOpen
  \bibfield  {author} {\bibinfo {author} {\bibfnamefont {H.}~\bibnamefont
  {Nemura}} \emph {et~al.},\ }in\ \href
  {https://inspirehep.net/record/1452812/files/arXiv:1604.08346.pdf} {\emph
  {\bibinfo {booktitle} {{12th International Conference on Hypernuclear and
  Strange Particle Physics (HYP 2015) Sendai, Japan, September 7-12, 2015}}}}\
  (\bibinfo {year} {2016})\ \Eprint {http://arxiv.org/abs/1604.08346}
  {arXiv:1604.08346 [hep-lat]} \BibitemShut {NoStop}%
\bibitem [{\citenamefont {Carbone}(2015)}]{CarboneHALprivcomm}%
  \BibitemOpen
  \bibfield  {author} {\bibinfo {author} {\bibfnamefont {A.}~\bibnamefont
  {Carbone}},\ }\href@noop {} {}\bibinfo {howpublished} {private communication}
  (\bibinfo {year} {2015})\BibitemShut {NoStop}%
\bibitem [{\citenamefont {Barbieri}\ and\ \citenamefont
  {Carbone}(2017)}]{Barbieri2017LNP}%
  \BibitemOpen
  \bibfield  {author} {\bibinfo {author} {\bibfnamefont {C.}~\bibnamefont
  {Barbieri}}\ and\ \bibinfo {author} {\bibfnamefont {A.}~\bibnamefont
  {Carbone}},\ }\enquote {\bibinfo {title} {Self-consistent green's function
  approaches},}\ in\ \href@noop {} {\emph {\bibinfo {booktitle} {An Advanced
  Course in Computational Nuclear Physics: Bridging the Scales from Quarks to
  Neutron Stars}}},\ \bibinfo {series} {Lect. Notes Phys.}, Vol.\ \bibinfo
  {volume} {936},\ \bibinfo {editor} {edited by\ \bibinfo {editor}
  {\bibfnamefont {M.}~\bibnamefont {Hjorth-Jensen}}, \bibinfo {editor}
  {\bibfnamefont {M.~P.}\ \bibnamefont {Lombardo}}, \ and\ \bibinfo {editor}
  {\bibfnamefont {U.}~\bibnamefont {van Kolck}}}\ (\bibinfo  {publisher}
  {Springer International Publishing},\ \bibinfo {address} {Cham},\ \bibinfo
  {year} {2017})\ Chap.~\bibinfo {chapter} {11}, pp.\ \bibinfo {pages}
  {571--644}\BibitemShut {NoStop}%
\bibitem [{\citenamefont {Fetter}\ and\ \citenamefont
  {Walecka}(2003)}]{fetter2003quantum}%
  \BibitemOpen
  \bibfield  {author} {\bibinfo {author} {\bibfnamefont {A.}~\bibnamefont
  {Fetter}}\ and\ \bibinfo {author} {\bibfnamefont {J.}~\bibnamefont
  {Walecka}},\ }\href {https://books.google.co.uk/books?id=0wekf1s83b0C} {\emph
  {\bibinfo {title} {Quantum Theory of Many-particle Systems}}},\ Dover Books
  on Physics\ (\bibinfo  {publisher} {Dover Publications},\ \bibinfo {year}
  {2003})\BibitemShut {NoStop}%
\bibitem [{\citenamefont {Dickhoff}\ and\ \citenamefont
  {Van~Neck}(2008)}]{dickhoff2008many}%
  \BibitemOpen
  \bibfield  {author} {\bibinfo {author} {\bibfnamefont {W.~H.}\ \bibnamefont
  {Dickhoff}}\ and\ \bibinfo {author} {\bibfnamefont {D.}~\bibnamefont
  {Van~Neck}},\ }\href@noop {} {\emph {\bibinfo {title} {Many-body theory
  exposed!}}},\ \bibinfo {edition} {2nd}\ ed.\ (\bibinfo  {publisher} {World
  Scientific Publishing, London},\ \bibinfo {year} {2008})\BibitemShut
  {NoStop}%
\bibitem [{\citenamefont {Schirmer}\ \emph {et~al.}(1983)\citenamefont
  {Schirmer}, \citenamefont {Cederbaum},\ and\ \citenamefont
  {Walter}}]{Schirmer1983ADCn}%
  \BibitemOpen
  \bibfield  {author} {\bibinfo {author} {\bibfnamefont {J.}~\bibnamefont
  {Schirmer}}, \bibinfo {author} {\bibfnamefont {L.~S.}\ \bibnamefont
  {Cederbaum}}, \ and\ \bibinfo {author} {\bibfnamefont {O.}~\bibnamefont
  {Walter}},\ }\href {\doibase 10.1103/PhysRevA.28.1237} {\bibfield  {journal}
  {\bibinfo  {journal} {Phys. Rev. A}\ }\textbf {\bibinfo {volume} {28}},\
  \bibinfo {pages} {1237} (\bibinfo {year} {1983})}\BibitemShut {NoStop}%
\bibitem [{\citenamefont {Barbieri}\ \emph {et~al.}(2007)\citenamefont
  {Barbieri}, \citenamefont {Van~Neck},\ and\ \citenamefont
  {Dickhoff}}]{Barbieri2007Atoms}%
  \BibitemOpen
  \bibfield  {author} {\bibinfo {author} {\bibfnamefont {C.}~\bibnamefont
  {Barbieri}}, \bibinfo {author} {\bibfnamefont {D.}~\bibnamefont {Van~Neck}},
  \ and\ \bibinfo {author} {\bibfnamefont {W.~H.}\ \bibnamefont {Dickhoff}},\
  }\href {\doibase 10.1103/PhysRevA.76.052503} {\bibfield  {journal} {\bibinfo
  {journal} {Phys. Rev. A}\ }\textbf {\bibinfo {volume} {76}},\ \bibinfo
  {pages} {052503} (\bibinfo {year} {2007})}\BibitemShut {NoStop}%
\bibitem [{\citenamefont {Som\`a}\ \emph
  {et~al.}(2014{\natexlab{b}})\citenamefont {Som\`a}, \citenamefont
  {Barbieri},\ and\ \citenamefont {Duguet}}]{Soma2014GkvII}%
  \BibitemOpen
  \bibfield  {author} {\bibinfo {author} {\bibfnamefont {V.}~\bibnamefont
  {Som\`a}}, \bibinfo {author} {\bibfnamefont {C.}~\bibnamefont {Barbieri}}, \
  and\ \bibinfo {author} {\bibfnamefont {T.}~\bibnamefont {Duguet}},\ }\href
  {\doibase 10.1103/PhysRevC.89.024323} {\bibfield  {journal} {\bibinfo
  {journal} {Phys. Rev. C}\ }\textbf {\bibinfo {volume} {89}},\ \bibinfo
  {pages} {024323} (\bibinfo {year} {2014}{\natexlab{b}})}\BibitemShut
  {NoStop}%
\bibitem [{\citenamefont {Hjorth-Jensen}\ \emph {et~al.}(1995)\citenamefont
  {Hjorth-Jensen}, \citenamefont {Kuo},\ and\ \citenamefont
  {Osnes}}]{HJORTHJENSEN1995125}%
  \BibitemOpen
  \bibfield  {author} {\bibinfo {author} {\bibfnamefont {M.}~\bibnamefont
  {Hjorth-Jensen}}, \bibinfo {author} {\bibfnamefont {T.~T.}\ \bibnamefont
  {Kuo}}, \ and\ \bibinfo {author} {\bibfnamefont {E.}~\bibnamefont {Osnes}},\
  }\href {\doibase http://dx.doi.org/10.1016/0370-1573(95)00012-6} {\bibfield
  {journal} {\bibinfo  {journal} {Physics Reports}\ }\textbf {\bibinfo {volume}
  {261}},\ \bibinfo {pages} {125 } (\bibinfo {year} {1995})}\BibitemShut
  {NoStop}%
\bibitem [{\citenamefont {Engeland}\ \emph {et~al.}()\citenamefont {Engeland},
  \citenamefont {Hjorth-Jensen},\ and\ \citenamefont
  {Jansen}}]{engeland2008cens}%
  \BibitemOpen
  \bibfield  {author} {\bibinfo {author} {\bibfnamefont {T.}~\bibnamefont
  {Engeland}}, \bibinfo {author} {\bibfnamefont {M.}~\bibnamefont
  {Hjorth-Jensen}}, \ and\ \bibinfo {author} {\bibfnamefont {G.}~\bibnamefont
  {Jansen}},\ }\href
  {https://github.com/ManyBodyPhysics/CENS/tree/master/MBPT/} {\bibinfo
  {journal} {{\em CENS, a Computational Environment for Nuclear Structure},
  https://github.com/ManyBodyPhysics/CENS/tree/master/MBPT/}\ }\BibitemShut
  {NoStop}%
\bibitem [{\citenamefont {Gad}\ and\ \citenamefont
  {M\"uther}(2002)}]{PhysRevC.66.044301}%
  \BibitemOpen
\bibfield  {journal} {  }\bibfield  {author} {\bibinfo {author} {\bibfnamefont
  {K.}~\bibnamefont {Gad}}\ and\ \bibinfo {author} {\bibfnamefont
  {H.}~\bibnamefont {M\"uther}},\ }\href {\doibase 10.1103/PhysRevC.66.044301}
  {\bibfield  {journal} {\bibinfo  {journal} {Phys. Rev. C}\ }\textbf {\bibinfo
  {volume} {66}},\ \bibinfo {pages} {044301} (\bibinfo {year}
  {2002})}\BibitemShut {NoStop}%
\bibitem [{\citenamefont {Luu}\ \emph {et~al.}(2004)\citenamefont {Luu},
  \citenamefont {Bogner}, \citenamefont {Haxton},\ and\ \citenamefont
  {Navr\'atil}}]{Luu2004BlchHor}%
  \BibitemOpen
  \bibfield  {author} {\bibinfo {author} {\bibfnamefont {T.~C.}\ \bibnamefont
  {Luu}}, \bibinfo {author} {\bibfnamefont {S.}~\bibnamefont {Bogner}},
  \bibinfo {author} {\bibfnamefont {W.~C.}\ \bibnamefont {Haxton}}, \ and\
  \bibinfo {author} {\bibfnamefont {P.}~\bibnamefont {Navr\'atil}},\ }\href
  {\doibase 10.1103/PhysRevC.70.014316} {\bibfield  {journal} {\bibinfo
  {journal} {Phys. Rev. C}\ }\textbf {\bibinfo {volume} {70}},\ \bibinfo
  {pages} {014316} (\bibinfo {year} {2004})}\BibitemShut {NoStop}%
\bibitem [{\citenamefont {Hergert}\ and\ \citenamefont
  {Roth}(2009)}]{Hergert2009Tkin}%
  \BibitemOpen
  \bibfield  {author} {\bibinfo {author} {\bibfnamefont {H.}~\bibnamefont
  {Hergert}}\ and\ \bibinfo {author} {\bibfnamefont {R.}~\bibnamefont {Roth}},\
  }\href {\doibase http://dx.doi.org/10.1016/j.physletb.2009.10.100} {\bibfield
   {journal} {\bibinfo  {journal} {Physics Letters B}\ }\textbf {\bibinfo
  {volume} {682}},\ \bibinfo {pages} {27 } (\bibinfo {year}
  {2009})}\BibitemShut {NoStop}%
\bibitem [{\citenamefont {More}\ \emph {et~al.}(2013)\citenamefont {More},
  \citenamefont {Ekstr\"om}, \citenamefont {Furnstahl}, \citenamefont {Hagen},\
  and\ \citenamefont {Papenbrock}}]{More2013InfraredExtr}%
  \BibitemOpen
  \bibfield  {author} {\bibinfo {author} {\bibfnamefont {S.~N.}\ \bibnamefont
  {More}}, \bibinfo {author} {\bibfnamefont {A.}~\bibnamefont {Ekstr\"om}},
  \bibinfo {author} {\bibfnamefont {R.~J.}\ \bibnamefont {Furnstahl}}, \bibinfo
  {author} {\bibfnamefont {G.}~\bibnamefont {Hagen}}, \ and\ \bibinfo {author}
  {\bibfnamefont {T.}~\bibnamefont {Papenbrock}},\ }\href {\doibase
  10.1103/PhysRevC.87.044326} {\bibfield  {journal} {\bibinfo  {journal} {Phys.
  Rev. C}\ }\textbf {\bibinfo {volume} {87}},\ \bibinfo {pages} {044326}
  (\bibinfo {year} {2013})}\BibitemShut {NoStop}%
\bibitem [{\citenamefont {Nemura}(2014)}]{Nemura2016WSPC_procs}%
  \BibitemOpen
  \bibfield  {author} {\bibinfo {author} {\bibfnamefont {H.}~\bibnamefont
  {Nemura}},\ }\href {\doibase 10.1142/S0218301314610060} {\bibfield  {journal}
  {\bibinfo  {journal} {Int. Jour. Mod. Phys. E}\ }\textbf {\bibinfo {volume}
  {23}},\ \bibinfo {pages} {1461006} (\bibinfo {year} {2014})}\BibitemShut
  {NoStop}%
\bibitem [{\citenamefont {Shavitt}\ and\ \citenamefont
  {Bartlett}(2009)}]{shavitt2009many}%
  \BibitemOpen
  \bibfield  {author} {\bibinfo {author} {\bibfnamefont {I.}~\bibnamefont
  {Shavitt}}\ and\ \bibinfo {author} {\bibfnamefont {R.}~\bibnamefont
  {Bartlett}},\ }\href {https://books.google.co.uk/books?id=SWw6ac1NHZYC}
  {\emph {\bibinfo {title} {Many-Body Methods in Chemistry and Physics: MBPT
  and Coupled-Cluster Theory}}},\ Cambridge Molecular Science\ (\bibinfo
  {publisher} {Cambridge University Press},\ \bibinfo {year}
  {2009})\BibitemShut {NoStop}%
\bibitem [{\citenamefont {Vries}\ \emph {et~al.}(1987)\citenamefont {Vries},
  \citenamefont {Jager},\ and\ \citenamefont {Vries}}]{DEVRIES1987chradii}%
  \BibitemOpen
  \bibfield  {author} {\bibinfo {author} {\bibfnamefont {H.~D.}\ \bibnamefont
  {Vries}}, \bibinfo {author} {\bibfnamefont {C.~D.}\ \bibnamefont {Jager}}, \
  and\ \bibinfo {author} {\bibfnamefont {C.~D.}\ \bibnamefont {Vries}},\ }\href
  {\doibase http://dx.doi.org/10.1016/0092-640X(87)90013-1} {\bibfield
  {journal} {\bibinfo  {journal} {Atomic Data and Nuclear Data Tables}\
  }\textbf {\bibinfo {volume} {36}},\ \bibinfo {pages} {495 } (\bibinfo {year}
  {1987})}\BibitemShut {NoStop}%
\bibitem [{\citenamefont {Angeli}\ and\ \citenamefont
  {Marinova}(2013)}]{Angeli2013chradii}%
  \BibitemOpen
  \bibfield  {author} {\bibinfo {author} {\bibfnamefont {I.}~\bibnamefont
  {Angeli}}\ and\ \bibinfo {author} {\bibfnamefont {K.}~\bibnamefont
  {Marinova}},\ }\href {\doibase http://dx.doi.org/10.1016/j.adt.2011.12.006}
  {\bibfield  {journal} {\bibinfo  {journal} {Atomic Data and Nuclear Data
  Tables}\ }\textbf {\bibinfo {volume} {99}},\ \bibinfo {pages} {69 } (\bibinfo
  {year} {2013})}\BibitemShut {NoStop}%
\bibitem [{\citenamefont {Cipollone}\ \emph {et~al.}(2015)\citenamefont
  {Cipollone}, \citenamefont {Barbieri},\ and\ \citenamefont
  {Navr\'atil}}]{Cipollone2015OxSpFnct}%
  \BibitemOpen
  \bibfield  {author} {\bibinfo {author} {\bibfnamefont {A.}~\bibnamefont
  {Cipollone}}, \bibinfo {author} {\bibfnamefont {C.}~\bibnamefont {Barbieri}},
  \ and\ \bibinfo {author} {\bibfnamefont {P.}~\bibnamefont {Navr\'atil}},\
  }\href {\doibase 10.1103/PhysRevC.92.014306} {\bibfield  {journal} {\bibinfo
  {journal} {Phys. Rev. C}\ }\textbf {\bibinfo {volume} {92}},\ \bibinfo
  {pages} {014306} (\bibinfo {year} {2015})}\BibitemShut {NoStop}%
\bibitem [{\citenamefont {M\"uther}\ and\ \citenamefont
  {Sick}(2004)}]{MuthSickPRC2004_rohr}%
  \BibitemOpen
  \bibfield  {author} {\bibinfo {author} {\bibfnamefont {H.}~\bibnamefont
  {M\"uther}}\ and\ \bibinfo {author} {\bibfnamefont {I.}~\bibnamefont
  {Sick}},\ }\href {\doibase 10.1103/PhysRevC.70.041301} {\bibfield  {journal}
  {\bibinfo  {journal} {Phys. Rev. C}\ }\textbf {\bibinfo {volume} {70}},\
  \bibinfo {pages} {041301} (\bibinfo {year} {2004})}\BibitemShut {NoStop}%
\bibitem [{\citenamefont {Murano}\ \emph {et~al.}(2014)\citenamefont {Murano},
  \citenamefont {Ishii}, \citenamefont {Aoki}, \citenamefont {Doi},
  \citenamefont {Hatsuda}, \citenamefont {Ikeda}, \citenamefont {Inoue},
  \citenamefont {Nemura},\ and\ \citenamefont {Sasaki}}]{Murano:2013xxa}%
  \BibitemOpen
  \bibfield  {author} {\bibinfo {author} {\bibfnamefont {K.}~\bibnamefont
  {Murano}}, \bibinfo {author} {\bibfnamefont {N.}~\bibnamefont {Ishii}},
  \bibinfo {author} {\bibfnamefont {S.}~\bibnamefont {Aoki}}, \bibinfo {author}
  {\bibfnamefont {T.}~\bibnamefont {Doi}}, \bibinfo {author} {\bibfnamefont
  {T.}~\bibnamefont {Hatsuda}}, \bibinfo {author} {\bibfnamefont
  {Y.}~\bibnamefont {Ikeda}}, \bibinfo {author} {\bibfnamefont
  {T.}~\bibnamefont {Inoue}}, \bibinfo {author} {\bibfnamefont
  {H.}~\bibnamefont {Nemura}}, \ and\ \bibinfo {author} {\bibfnamefont
  {K.}~\bibnamefont {Sasaki}},\ }\href {\doibase
  http://dx.doi.org/10.1016/j.physletb.2014.05.061} {\bibfield  {journal}
  {\bibinfo  {journal} {Physics Letters B}\ }\textbf {\bibinfo {volume}
  {735}},\ \bibinfo {pages} {19 } (\bibinfo {year} {2014})}\BibitemShut
  {NoStop}%
\end{thebibliography}%

\end{document}